\documentclass[aip,jcp,reprint]{revtex4-1}
\newcommand\T{\rule{0pt}{3.0ex}}       % Top strut

\usepackage{graphicx}
\usepackage{amsmath}
\usepackage{longtable}
\usepackage{pbox}
\usepackage[breaklinks=true,colorlinks,citecolor=blue,linkcolor=blue,urlcolor=blue]{hyperref}
\begin{document}
\title{A   study of accurate exchange-correlation functionals through adiabatic connection}

% author. Explanatory text should go in the []'s, actual e-mailcitations.bib   Makefile                    thesis.aux      thesis.fdb_latexmk  thesis.lot  thesis.tex


\author{Rabeet Singh}
\email[]{rabeetsingh@gmail.com}
%\thanks{}
%\altaffiliation{Department of Physics, Indian Institute of Technology, Kanpur-208016}https://webmail2.iitk.ac.in/webmail/src/signout.php
\affiliation{Department of Physics, Indian Institute of Technology Kanpur, Kanpur-208016, India}
\author{Manoj K. Harbola}
\email[]{mkh@iitk.ac.in}
%\thanks{}
\affiliation{Department of Physics, Indian Institute of Technology Kanpur, Kanpur-208016, India}
\date{\today}
\begin{abstract}
A systematic way of improving exchange-correlation energy functionals of density functional theory has been to make them satisfy more and more exact relations. Starting from the initial GGA functionals, this has culminated into the recently proposed SCAN(Strongly constrained and appropriately normed) functional that satisfies several known constraints and is appropriately normed. The ultimate test for the functionals developed is the accuracy of energy calculated by employing them. In this paper, we test these exchange-correlation functionals $-$the GGA hybrid functionals B3LYP and PBE0, and the meta-GGA functional SCAN$-$ from a different perspective. We study how accurately these functionals reproduce the exchange-correlation energy when electron-electron interaction is scaled as $\alpha V_{ee}$ with $\alpha$ varying between $0$ and $1$. Our study reveals interesting comparison between these functionals and the associated difference $T_c$ between the interacting and the non-interacting kinetic energy for the same density. 
\end{abstract}
%\keywords{}
\maketitle
% body of paper here - Use proper section commandshttps://webmail2.iitk.ac.in/webmail/src/signout.php
\section{Introduction}
Since the inception of density functional theory\cite{Hohn,Yang}(DFT) and its first demonstration\cite{Tong} of the solution of Kohn-Sham equation\cite{Kohn} using the LDA\cite{Yang}, accuracy of the exchange-correlation(xc) functionals has increased manyfold.  Efforts in understanding and improving the xc functionals have been made by looking at exact properties satisfied by these functionals or associated quantities. Thus the accuracy of the LDA was first explained\cite{Gun1} in terms of the corresponding exchange-correlation hole satisfying the charge neutrality condition. An attempt to go beyond the LDA for exchange by incorporating gradient correction was first made\cite{Herman} on dimensional basis and gave the gradient expansion approximation(GEA) for exchange. The expansion was formally derived\cite{Sham1971} by using the response formula of an electron gas.  However, the corresponding Fermi hole\cite{Gross} is unphysical\cite{Perd85}.  Correcting deficiency of the GEA led\cite{Perd85} to the first generalized gradient approximation(GGA) for the exchange functional. Along similar line of thinking another GGA functional was proposed\cite{Bec88} that was based on the exact asymptotic dependence of exchange energy density far away from a system. 
\par For the correlation functionals, the LDA was first given by Wigner\cite{Wigner1,Wigner2,Wigner3}. It was later derived for high density of electrons by Gelmann and Brueckner\cite{Gelbru} and gradient correction was given by Ma and Brueckner\cite{Mabru}. Accurate functionals were later constructed by parameterizing the Monte-Carlo calculations of Ceperley and Alder\cite{Cepal} and making sure that the high density limit of the correlation functional was reproduced. Further work on correlation functionals is that given by Langreth-Mehl\cite{Lanme}, Perdew-Langreth\cite{Lanpr}, Perdew and Wang\cite{Perwa} and the Perdew-Burke \& Ernzerhof\cite{GGA}(PBE).     
\par As is clear, the approach to developing functionals has been to start with the simplest functional given by the LDA, making gradient correction to it up the second order or the fourth order\cite{Sven,PSBarth} and then demanding that the resulting functional satisfy exact conditions for them. The results have been that highly accurate energies can now be calculated using density functionals developed over the past two decades. 
\par Two highly accurate hybrid functionals\cite{Beck} employing the Becke and the PBE functionals, respectively, are the B3LYP\cite{Becke,LYP} and PBE0\cite{GGA,Perd,PBE0} functionals. While the Becke functional used in B3LYP satisfies the exact asymptotic dependence of the exchange energy density, PBE functional satisfies 11 known relations. Recently a new functional SCAN\cite{Scan}(strongly constrained and appropriately normed) has been  proposed that satisfies 17 known properties and is appropriately normed. These three functionals are the focus of our study in this paper.
\par As accuracy of the functionals increases, so do the expectations from density functional theory to describe large range of phenomenon and provide other quantities also with the same accuracy as the energy. In this context, it has recently been noted that the densities themselves are not as accurate as the energies given by DFT\cite{Medve}. Keeping this in mind, it is imperative that the investigations are done in testing the energy functionals from various perspectives and for different systems\cite{TruhF}. The work presented in this paper is our attempt in that direction. Here we test the functionals for simple spherical systems (atoms and jellium spheres that represent metallic clusters\cite{Knight1,Matt,Walt}) via adiabatic connection\cite{Jones,Langp1,Gun1,Gun2,Langp2} as the electron-electron interaction is varied for a system keeping its density unchanged. We compare the resulting curves with the exact results for two-electron systems and show how different functionals differ from the exact results. We then calculate difference $T_c$ in the kinetic energies of the true systems and the corresponding Kohn-Sham systems and show that it varies over a range of values for different systems and functionals. We then extend our study to larger systems using their Hartree-Fock density. We note that importance of adiabatic connection curves to aid progress in construction of energy functional was explored\cite{Frydb} in a study of such curves for two electron systems. 
\par In the following, we begin with a description of adiabatic connection and its relationship\cite{LP85} with the scaling relations. 
\begin{table*}
	\caption{Exchange and correlation energies (in atomic units) obtained from B3LYP, PBE0 and SCAN functionals using nearly exact densities for H$^-$ and He and the exact density for the Hookium atom. Results using Hartree-Fock densities for all the systems are also given.}
	\begin{ruledtabular}
		\begin{tabular}{cccccccccc}
			Atom& input density    &\multicolumn{4}{c}{Exchange Energy}  &\multicolumn{4}{c}{Correlation Energy} \\ 
			\cline{3-6} \cline{7-10} \T
			&            & Exact &  B3LYP & PBE0   & SCAN  &  Exact &  B3LYP & PBE0    & SCAN   \\ \hline \T 
			H$^-$       &Near exact  & \pbox{10cm}{-0.3809\cite{Cyru}\\ -0.3828\cite{Teal2}}& -0.3877&-0.3889&-0.3923&\pbox{10cm}{-0.0420\cite{Cyru}\\-0.0418\cite{Teal2}}& -0.0378&-0.0305 &-0.0298  \\
			&Hartree-Fock&       & -0.3943&-0.3947&-0.4024&        & -0.0391&-0.0327 &-0.0307 \\ \T \T 
			He          &Near Exact  &\pbox{10cm}{-1.0246\cite{Cyru}\\-1.0246\cite{Teal2}}& -1.0136&-1.0160&-1.0299&\pbox{10cm}{-0.0421\cite{Cyru}\\ -0.0422\cite{Teal2}} & -0.0569&-0.0419 &-0.0379  \\
			&Hartree-Fock&       & -1.0140&-1.0163&-1.0306&        & -0.0569&-0.0420 &-0.0379 \\ \T \T
			Hookium      &   Exact    &-0.5160\cite{Kais}& -0.4998&-0.4986&-0.5141& -0.0393\cite{Kais}& -0.0447&-0.0514 &-0.0355 \\
			&Hartree-Fock&       &  -0.4995 &-0.4984 & -0.5137       &   & -0.0447 & -0.0512 & -0.0355 \\ \hline 
      \multicolumn{2}{c}{Mean absolute percentage error}      &  &2.0 & 2.1 &  1.3       &   & 19.6 & 19.5 & 16.2
		\end{tabular}  
	\end{ruledtabular}
	\label{ex_ec_exd}
\end{table*}
\begin{table*}%[H] add [H] placement to break table across pages
	\caption{The difference $T_c$ between the interacting and noninteracting kinetic energies (in atomic units) for H$^-$, He and Hookium atom densities. Also shown in parenthesis after each approximate value is its percentage deviation from the exact value. Results obtained from Hartree-Fock densities are also shown.}
	\begin{ruledtabular}
		\begin{tabular}{cccccc}
			Atom	& input density &\multicolumn{4}{c}{$T_c$}     \\ 
			\cline{3-6} \T
			&    & Exact  & B3LYP     & PBE0       & SCAN  \\  \hline \T
			H$^-$ 	&Near exact   & 0.0279\cite{Teal2} &0.0199(28.7) & 0.0241(13.6)& 0.0217(22.2)        \\
			        &Hartree-Fock &                    &0.0208(25.4) & 0.0256(8.3) & 0.0224(19.7)         \\ \T \T
			He	&Near exact & 0.0367\cite{Teal2}   &0.0402(9.5) & 0.0376(2.4) & 0.0325(11.4)        \\
			        &Hartree-Fock &                    &0.0403(9.8) & 0.0377(2.7) & 0.0325(11.4)            \\ \T \T
			Hookium	&Exact & 0.0273\cite{Kais}         &0.0301(10.2) & 0.0355(30.0)& 0.0276(1.1)        \\
			        &Hartree-Fock &        &           0.0300(9.9)  & 0.0355(30.0)& 0.0275(0.7) \\ \hline 
            \multicolumn{2}{c}{Mean absolute percentage error}  &        &   16.1                &  15.3                 &  11.6
		\end{tabular}  
	\end{ruledtabular}
	\label{tc_exd}
\end{table*}
\begin{figure*}
	\caption{Exact $W_c^{\alpha}$ for H$^-$, He and the Hookium as a function of $\alpha$ and its comparison with $W_c^{\alpha}$ for the B3LYP, PBE0 and SCAN functionals \label{wxc_exd}}
	\includegraphics[scale=0.6]{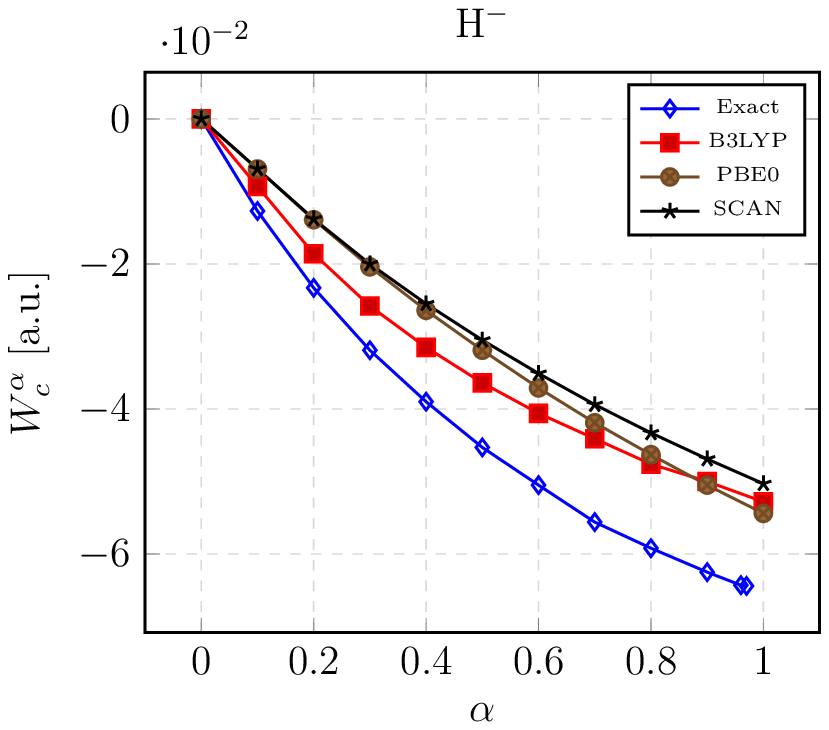} \hspace{0.5cm}%
	\includegraphics[scale=0.6]{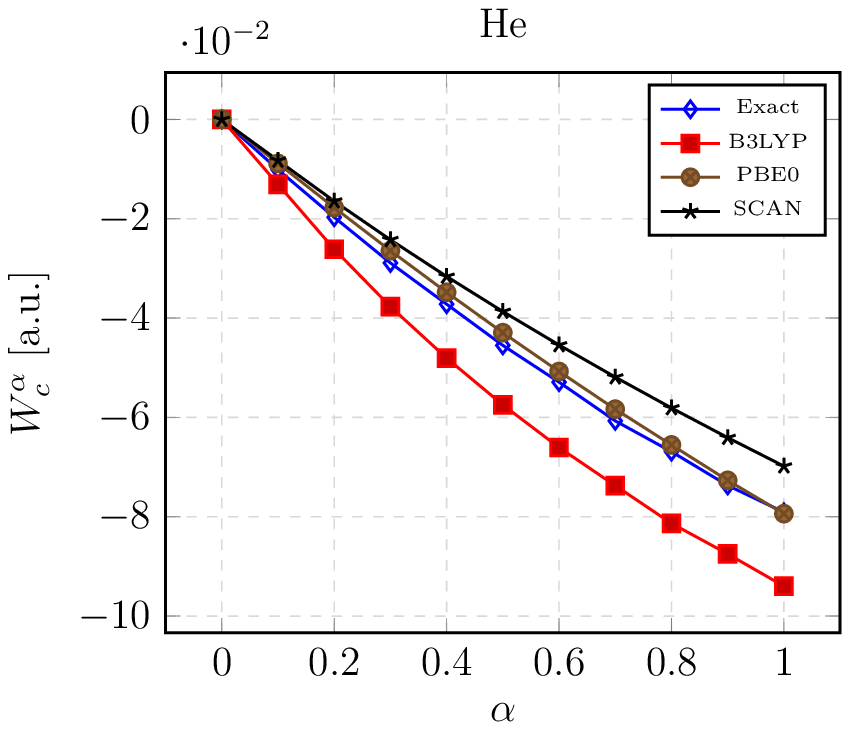}\hspace{0.5cm}
	\includegraphics[scale=0.6]{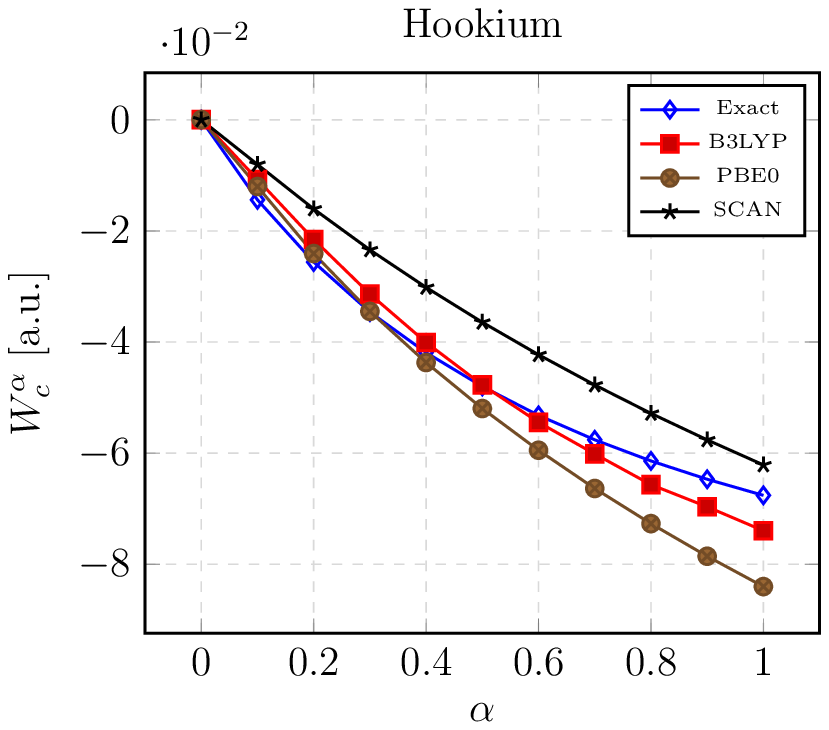}
\end{figure*}
\section{Adiabatic connection and scaling relations}
Exchange-correlation energy for a ground state density $\rho(\vec{r})$ in density functional theory can be defined\cite{Jones,Langp1,Gun1,Gun2,Langp2} in terms of the many-body wavefunction as
\begin{equation}
E_{xc}^{DFT}[\rho]=\int_{0}^{1} W_{xc}^{\alpha} d\alpha \label{adcn}
\end{equation}  
with $W_{xc}^{\alpha}=\langle \Psi^{\alpha} |V_{ee}|\Psi^{\alpha}\rangle -\frac{1}{2}  \iint \frac{\rho(\vec{r})\rho(\vec{r}')}{|\vec{r}-\vec{r}'|} d\vec{r} d\vec{r}'$. Here $V_{ee}$ is the electron-electron interaction energy and $\Psi^{\alpha}$ is the many-electron wavefunction that gives the density $\rho(\vec{r})$ and is the ground state solution of the Schr$\ddot{\text{o}}$dinger equation
\begin{equation}
[\hat{T} + V_{ext}^{\alpha} + \alpha V_{ee}]\Psi^{\alpha}=E^{\alpha}\Psi^{\alpha}, \label{scal}
\end{equation}
where $V_{ext}^{\alpha}$ is the external potential corresponding to the scaled Coulombic interaction $\alpha V_{ee}$. Notice that the traditional quantum-chemical exchange-correlation energy is $W_{xc}^{\alpha=1}$. The quantity\cite{Ernz}
\begin{equation}
T_c = E_{xc}^{DFT}-W_{xc}^{\alpha=1} \label{kico}
\end{equation}
is the difference between the exact kinetic energy of the true system and the non-interacting kinetic energy given by the corresponding Kohn-Sham equation. Eq. (\ref{adcn}) has been used to develop hybrid functionals\cite{Beck} that mix $E_{xc}^0[\rho]$, which is the exact exchange-energy calculated from the Kohn-Sham orbitals, and an approximation to $W_{xc}^{\alpha=1}[\rho]$ which is made by replacing it by $E_{xc}^{DFT}[\rho]$(approximate). Generalizing this approach, attempts have been made\cite{Ernz,Colon,Cohen,Alexe,Teal3,Neil} to model $W_{xc}^{\alpha}$ and integrate it to obtain an approximate $E_{xc}[\rho]$. In this paper, we do the opposite of this in that we calculate $W_{xc}^{\alpha}$ corresponding to an approximate exchange-correlation functional and compare this with the exact $W_{xc}^{\alpha}$ for systems where the latter can be calculated. This gives a different perspective in understanding an approximate functional. In addition, using Eq. (\ref{kico}) we also calculate $T_c$ corresponding to a given density for an approximate functional to learn about its contribution to the errors in the functional.
\begin{table*}
	\caption{Exchange and correlation energies (in atomic units) for atoms calculated using their HF densities for the functionals B3LYP, PBE0 and SCAN.}
	\begin{ruledtabular}
		\begin{tabular}{ccccccccc}
			Atom  &\multicolumn{4}{c}{Exchange Energy} &\multicolumn{4}{c}{Correlation Energy} \\ 
			\cline{2-5} \cline{6-9} \T
	  & Exact               & B3LYP   & PBE0     &  SCAN   & Exact                   & B3LYP  & PBE0      & SCAN   \\ \hline \T
	Be&-2.6735\cite{Teal3}  &-2.6332  & -2.6450 & -2.6603 &-0.0945\cite{Teal3}      & -0.1193&  -0.0855 & -0.0827 \\
	Ne&-12.1080\cite{Bec88} &-12.0435 & -12.0770& -12.1650&-0.3910\cite{Shubh,Shane}      & -0.4524&  -0.3513 & -0.3448 \\
	Ar&-30.1880\cite{Bec88} &-29.9767 & -30.0436& -30.2700&-0.7230\cite{Shubh,Shane}& -0.8799&  -0.7067 & -0.6905 \\
	Kr&-93.8900\cite{Bec88} &-93.4538 & -93.5397& -94.1067&-1.8500\cite{Shane}      & -2.0402&  -1.7672 & -1.7560 \\
	Xe&-179.2000\cite{Bec88}&-178.3916&-178.4799&-179.3922&-3.0000\cite{Shane}      & -3.2103&  -2.9182 & -2.8997 \\ \hline
  \multicolumn{2}{c}{Mean absolute percentage error}      &0.7   & 0.5 &  0.3       &   & 16.2 & 5.8 & 7.4
		\end{tabular} 
	\end{ruledtabular}
	\label{ex_ec_hf_in}
\end{table*}

\begin{table*}%[H] add [H] placement to break table across pages
	\caption{Exchange-correlation energies (in atomic units) for atoms. For H$^-$, He and the Hookium, the 
                 results for density functionals have been calculated using exact densities while for other systems,
                 their HF densities has been used. It is clear that the SCAN functional provides best cancellation of
                 errors between the exchange and correlation.}
	\begin{ruledtabular}
		\begin{tabular}{ccccc}
			Atom	&\multicolumn{4}{c}{Exchange-correlation energy}     \\ 
			\cline{2-5} \T
			     & Exact                & B3LYP    & PBE0      & SCAN  \\  \hline \T
			H$^-$&   -0.4229 &-0.4255   &-0.4194   & -0.4221       \\
			He   &   -1.0667 &-1.0705   &-1.0579   & -1.0678       \\
	              Hookium&   -0.5553 &-0.5445   &-0.5500   & -0.5496      \\
                      Be     &  -2.7680  &  -2.7525 &  -2.7305 &  -2.7430 \\
                      Ne     & -12.4990  & -12.4959 & -12.4283 & -12.5098 \\
                      Ar     & -30.9110  & -30.8566 & -30.7503 & -30.9605 \\
                      Kr     & -95.7400  & -95.4940 & -95.3069 & -95.8627 \\
                      Xe     &-182.2000  &-181.6019 &-181.3981 &-182.2919 \\ \hline 
                \multicolumn{2}{c}{Mean absolute percentage error}         &0.9    &1.2   &0.5
		\end{tabular}  
	\end{ruledtabular}
	\label{exc_atoms}
\end{table*}

\begin{table*} %add [H] placement to break table across pages
	\caption{The difference $T_c$ between the exact and Kohn-Sham kinetic energies (in atomic units) calculated using the Hartree-Fock densities.}
	\begin{ruledtabular}
		\begin{tabular}{ccccc}
			Atom    &\multicolumn{4}{c}{$T_c$}    \\ 
			\cline{2-5}  \T
			& Exact  & B3LYP    & PBE0 & SCAN \\  \hline \T
			Be	& 0.0725\cite{Teal3} & 0.0757 & 0.0717& 0.0666      \\ 
			Ne	& 0.3154\cite{Teal3} &0.3269 & 0.3073& 0.2941      \\ 
			Ar	&   $-$  &0.6517    & 0.6155   & 0.5941        \\
			Kr  &   $-$  &1.5943    & 1.5640   & 1.5243     \\ 
			Xe  &   $-$  &2.5678    & 2.5989   & 2.5453        
		\end{tabular}
	\end{ruledtabular}
	\label{tc_hf_in}
\end{table*}
	\begin{figure*}
		\caption{$W_c^{\alpha}$ for Be, Ne and Ar as a function of $\alpha$ obtained using Hartree-Fock densities in the B3LYP, PBE0 and SCAN functionals.\label{wxc_hf_in}}
		\includegraphics[scale=0.55]{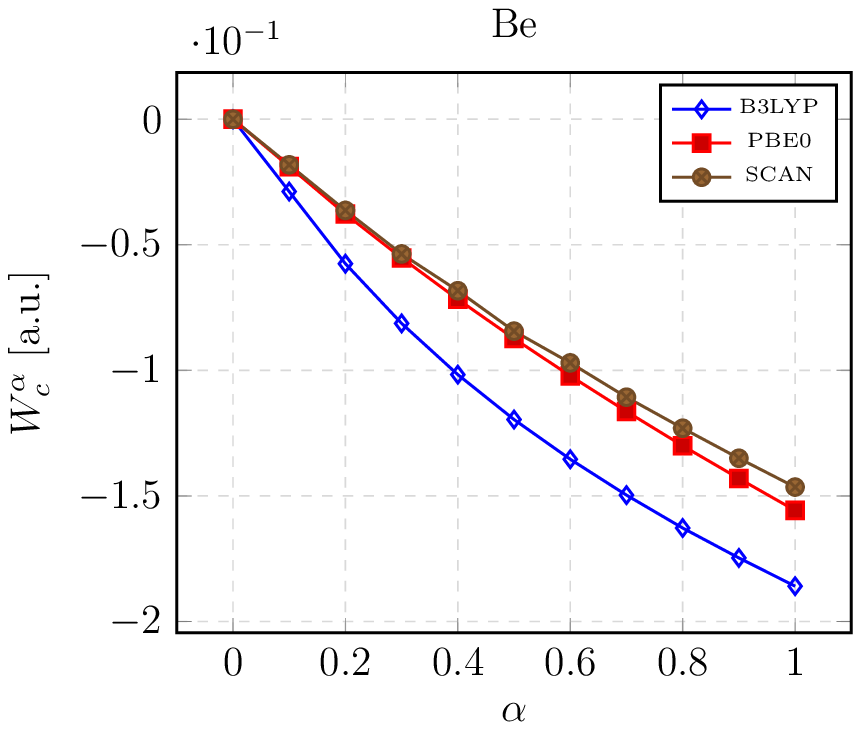} \hspace{0.5cm}%
		\includegraphics[scale=0.55]{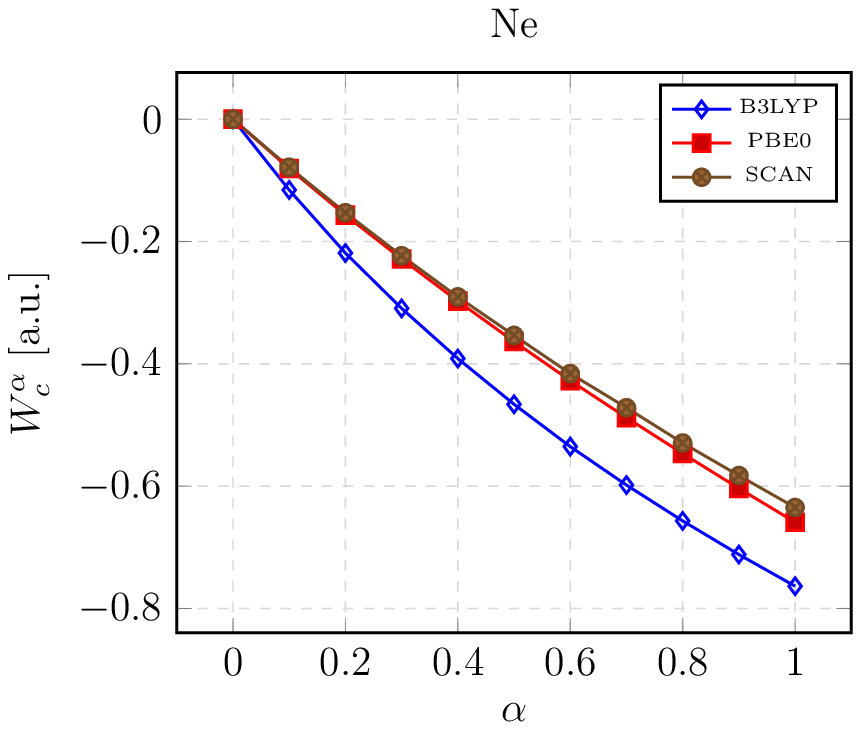}\hspace{0.5cm}
		\includegraphics[scale=0.55]{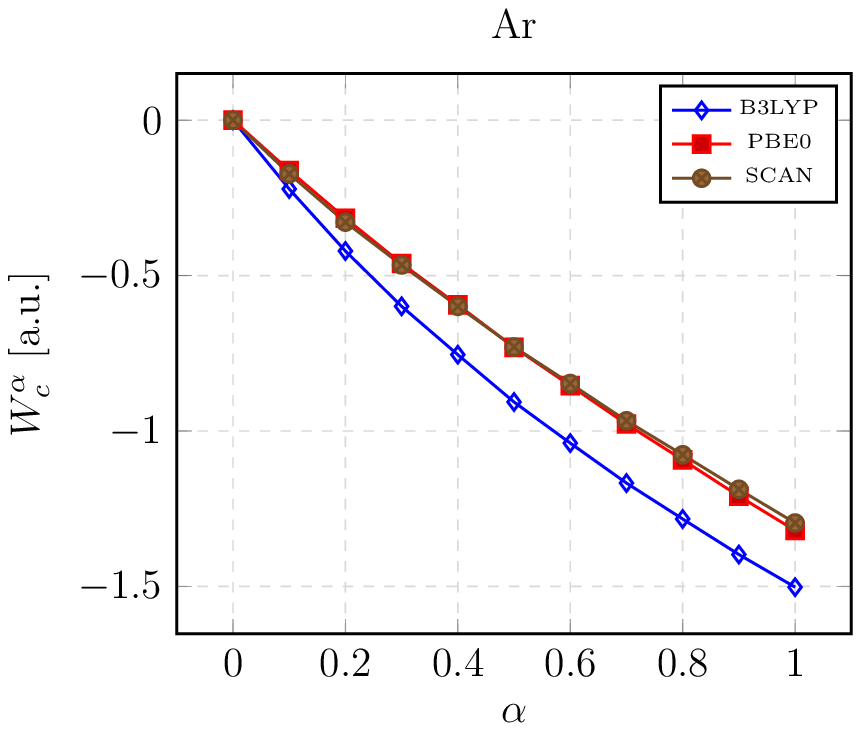}
	\end{figure*}
\subsection{Obtaining $W_{xc}^{\alpha}$ for $E_{xc}^{approx}[\rho]$}
For an approximate exchange-correlation energy functional $E_{xc}^{approx}[\rho]$, the corresponding\cite{LP85,Cohen,Alexe}
\begin{equation}
W_{xc}^{\alpha}[\rho] =\frac{d}{d\alpha} \Big(\alpha^2 E_{xc}[\rho_{\frac{1}{\alpha}}]\Big),  \label{wxca}
\end{equation}
where $\rho_{\frac{1}{\alpha}}$ is the scaled density given as 
\begin{equation}
\rho_{\frac{1}{\alpha}}(\vec{r}) = \frac{1}{\alpha^3} \rho\Big(\frac{\vec{r}}{\alpha}\Big).  \label{scde}
\end{equation}
For completeness, we give the derivation of Eq. (\ref{wxca}) in the appendix. Before proceeding with our calculations, we reflect on what could  the quantities $W_{xc}^{\alpha}$ in general, and $W_{xc}^{\alpha=0}$ and $W_{xc}^{\alpha=1}$ in particular, mean when an approximate exchange-correlation energy functional $E_{xc}^{approx}[\rho]$ is used in right hand side of Eq. (\ref{wxca}). A possible interpretation is as follows.
\par An approximate functional may be thought of being the exact functional for an inter-electron interaction $w(\vec{r}, \vec{r}')$ which could be different from the Coulombic interaction. However it should have the same dimensional dependence on $|\vec{r}-\vec{r}'|$ as the Coulomb interaction in order that the relations used in this paper can be applied to it. The density $\rho(\vec{r})$ employed in our study is that produced by solving the many-electron Schr$\ddot{\text{o}}$dinger equation with an external potential $\tilde{v}_{ext}(\vec{r})$ which is also different from the true external potential.  $W_{xc}^{\alpha=1}$ is then given as 
\begin{equation}
W_{xc}^{\alpha=1}[\rho]=\langle \tilde{\Psi} |\hat{W}|\tilde{\Psi} \rangle 
-\frac{1}{2} \iint w(\vec{r}, \vec{r}')\rho(\vec{r})\rho(\vec{r}') d\vec{r}d\vec{r}' \label{wxca1}
\end{equation}
where $\hat{W}=\frac{1}{2}\sum_{ij} w(\vec{r}_i, \vec{r}_j)$ and $\tilde{\Psi}$ is the solution of the Schr$\ddot{\text{o}}$dinger equation for the Hamiltonian with the modified external potential $\tilde{v}_{ext}(\vec{r})$ and the interaction $w(\vec{r}, \vec{r}')$. By constrained search, $\tilde{\Psi}$ is the wavefunction that gives the density $\rho(\vec{r})$ and minimizes the expectation value $\langle \hat{T}+\hat{W}\rangle $. The corresponding $W_{xc}^{\alpha}$ is then
\begin{equation}
W_{xc}^{\alpha}=\langle \tilde{\Psi}^{\alpha}|\hat{W}| \tilde{\Psi}^{\alpha} \rangle -\frac{1}{2} \iint \rho(\vec{r})\rho(\vec{r}') w(\vec{r}, \vec{r}') d\vec{r} d\vec{r}'
\end{equation} 
where $\tilde{\Psi}^{\alpha}$ gives the density $\rho(\vec{r})$ and minimizes the expectation value $\langle \tilde{\Psi}^{\alpha}| \hat{T} + \alpha \hat{W}| \tilde{\Psi}^{\alpha} \rangle$. With this understanding, we can obtain the approximate exchange energy $W_{x}^{approx}[\rho]$ and the kinetic component $T_c^{approx}$ for a given exchange-correlation functional as
\begin{equation*}
W_{x}^{approx}[\rho]=W_{xc}^{\alpha=0},
\end{equation*}
\begin{equation}
T_c^{approx}  =E_{xc}^{approx}[\rho]-W_{xc}^{\alpha=1}.\label{sc_tc}
\end{equation}
For a given approximate functional $E_{xc}^{approx}[\rho]$ and a given density $\rho(\vec{r})$, the quantity $W_{xc}^{\alpha}$ can be calculated using Eq. (\ref{wxca}). Note that the value of $W_{xc}^{\alpha=0}$ gives the exchange energy for the given functional while the $W_{xc}^{\alpha=1}$ is given by Eq.(\ref{wxca1}). Thus the corresponding kinetic energy difference $T_c^{approx}$ between the fully-interacting system and the associated Kohn-Sham system is easily calculated using Eq. (\ref{sc_tc}), where $E_{xc}^{approx}[\rho]$ is given directly from the functional and $W_{xc}^{\alpha=1}$ is calculated from Eq. (\ref{wxca}). 
	\begin{table*}
		\caption{Exchange and correlation energies per electron (in atomic units) of neutral jellium spheres using the Hartree-Fock density in the functionals B3LYP, PBE0 and SCAN. Exact values of exchange and correlation energies per electron are also given. The bulk value for exchange energy per electron is $-0.1145$[a.u.] and that for correlation is $-0.0318$\cite{Vwn, Teepa}.}
		\begin{ruledtabular}
			\begin{tabular}{ccccccccc}
				N &\multicolumn{4}{c}{Exchange Energy} &\multicolumn{4}{c}{Correlation Energy}  \\ 
				\cline{2-5} \cline{6-9} \T
				   & Exact  & B3LYP &  PBE0 & SCAN  & Exact\cite{Sottile_PRB.64.045105}  &  B3LYP &  PBE0   &  SCAN    \\ \hline \T
		                2  &-0.1093 &-0.1059&-0.1059&-0.1087&-0.0143 &-0.0150 & -0.0180 & -0.0136   \\
				8  &-0.1078 &-0.1063&-0.1063&-0.1086&-0.0207 &-0.0164 & -0.0231 & -0.0198   \\
				18 &-0.1093 &-0.1085&-0.1085&-0.1098&-0.0236 &-0.0171 & -0.0251 & -0.0228   \\
				20 &-0.1077 &-0.1071&-0.1071&-0.1083&-0.0241 &-0.0171 & -0.0254 & -0.0235   \\
				34 &-0.1103 &-0.1097&-0.1097&-0.1106&-0.0250 &-0.0175 & -0.0265 & -0.0246   \\
				40 &-0.1079 &-0.1077&-0.1077&-0.1081&-0.0259 &-0.0175 & -0.0268 & -0.0257  \\
				58 &-0.1107 &-0.1102&-0.1102&-0.1110&-0.0261 &-0.0178 & -0.0274 & -0.0259 \\ \hline 
  \multicolumn{2}{c}{Mean absolute percentage error}& 1.0& 1.0& 0.4 &  &25.2 &9.1& 2.6
			\end{tabular}
		\end{ruledtabular}
		\label{ex_ec_hf_cl}
	\end{table*}

\begin{table*}%[H] add [H] placement to break table across pages
	\caption{Exchange-correlation energies (in atomic units) per electron for neutral jellium spheres using their
                 Hartree-Fock densities in the B3LYP, PBE0 and SCAN functionals in comparision to the exact values.
                 The Bulk value for the exchange-correlation energy per electron is $-0.1463$[a.u.].}
	\begin{ruledtabular}
		\begin{tabular}{ccccc}
			N	&\multicolumn{4}{c}{Exchange-correlation energy}     \\ 
			\cline{2-5} \T
			     & Exact                & B3LYP    & PBE0      & SCAN  \\  \hline \T
		 2   &-0.1236&-0.1209&-0.1239&-0.1223 \\         
		 8   &-0.1285&-0.1227&-0.1294&-0.1284 \\
		 18  &-0.1329&-0.1256&-0.1336&-0.1326 \\
		 20  &-0.1318&-0.1242&-0.1325&-0.1318 \\
		 34  &-0.1353&-0.1272&-0.1362&-0.1352 \\
		 40  &-0.1338&-0.1252&-0.1345&-0.1338 \\
		 58  &-0.1368&-0.1280&-0.1376&-0.1369 \\  \hline 
                \multicolumn{2}{c}{Mean absolute percentage error} & 5.3&0.5&0.2
		\end{tabular}  
	\end{ruledtabular}
	\label{exc_jc}
\end{table*}

	\begin{table}
		\caption{The difference $T_c$ per electron between the exact and Kohn-Sham kinetic energies (atomic units) calculated using the Hartree-Fock densities. By the virial relation\cite{Wang} $t_c=0.0082$au for the bulk.}
		\begin{ruledtabular}
			\begin{tabular}{cccc}
				N     &\multicolumn{3}{c}{$T_c$}    \\  
				\cline{2-4} \T
				         & B3LYP & PBE0   &   SCAN \\  \hline \T
				2 &0.0063 &0.0102 & 0.0085        \\
				8 &0.0057 &0.0120 & 0.0110         \\
				18&0.0056 &0.0127 & 0.0122         \\
				20&0.0055 &0.0127 & 0.0122        \\
				34&0.0056 &0.0132 & 0.0128         \\
				40&0.0054 &0.0132 & 0.0129        \\
				58&0.0055 &0.0135 & 0.0132        
			\end{tabular}
		\end{ruledtabular}
		\label{tc_hf_cl}
	\end{table}	
\section{Results}
In this section we begin by presenting results for the total exchange-correlation energies and $W_{xc}^{\alpha}$ for the functionals the B3LYP\cite{Becke,LYP} and the PBE0\cite{GGA}, and for the most recently proposed\cite{Scan} SCAN functional. We first perform our study using the exact densities for two-electron systems of the He atom and H$^-$ ion and the Hookium atom\cite{Rabi,Taut,Kais}.  This enables us to analyze the functionals for two different kinds of external potentials : The external potential for H$^-$ and He atom has $-\frac{1}{r}$ dependence whereas the external potential in the Hookium atom has $r^2$ dependence where $r$ is the distance from the origin. After this, we extend our study to larger systems using Hartree-Fock densities. We show that the exchange energies obtained from different functionals are close to each other and the difference between exchange-correlation energies arises mainly from the correlation energy difference. This is further reflected in the numbers for the corresponding $T_c$.

\subsection{Results for exact densities}
Shown in Table \ref{ex_ec_exd} are the results for the exchange and correlation energies for the He, H$^-$ and the Hookium atom. These energies have been calculated from the near exact semi analytic densities\cite{Rabi,Rabi2} for H$^-$ and He and the exact density\cite{Kais,Taut} for the Hookium atom. We note that SCAN functional has also been studied\cite{Sun_16} for Hooke's atom recently. It is observed from the Table that the exchange energy for all the systems comes out to be roughly the same for the three functionals. However, the SCAN functional gives slightly larger magnitude than the other two functionals. On the other hand, the correlation energy varies by a relatively larger amount among the three functionals with its magnitude being the smallest for the SCAN functional. 
\par Next we employ Eq. (\ref{wxca}) to obtain $W_{xc}^{\alpha}$ curves for the functionals being studied and compare them with the exact curves\cite{Rabi2}. Since the exact part $W_x^{\alpha}$ of it is a constant equal to the exchange energy, we display only $W_c^{\alpha}$ in Fig. \ref{wxc_exd} where we have plotted $W_c^{\alpha}$ against $\alpha$ for the H$^-$, He and the Hookium atom. From the figure, it is evident that for all the systems the curves are close to each other. 
\par We now calculate $T_c$ for these systems and compare the values obtained from different functionals with the exact results. The values of $T_c$ for density functionals are obtained by using Eq. (\ref{sc_tc}). These are displayed in Table \ref{tc_exd}. It is evident from Table \ref{tc_exd} that $T_c$ can vary quite a bit from functional to functional. Although we have also given the percentage errors for each functional, these are not very meaningful because the exact numbers are quite small. We note that $T_c$ given by the PBE0 functional is closest to the exact values for H$^-$ and He while SCAN functional is nearly exact for the Hookium atom. 
\par Having studied the functionals for small systems where the exact $W_{xc}^{\alpha}$ curve could be obtained easily, we now extend our study to larger systems using their Hartree-Fock or Hartree-Fock like densities. In this study, we are not able to calculate the exact $W_c^{\alpha}$ so the study is limited to a comparison of $W_c^{\alpha}$ curves for different functionals. This itself is significant because it brings out the differences between the three functionals.
 \begin{figure*}
 	\caption{$W_c^{\alpha}$ for six different clusters as a function of $\alpha$ obtained using Hatree-Fock densities in the B3LYP, PBE0, and SCAN functionals.\label{wxc_hf_cl}}
 	\includegraphics[scale=0.55]{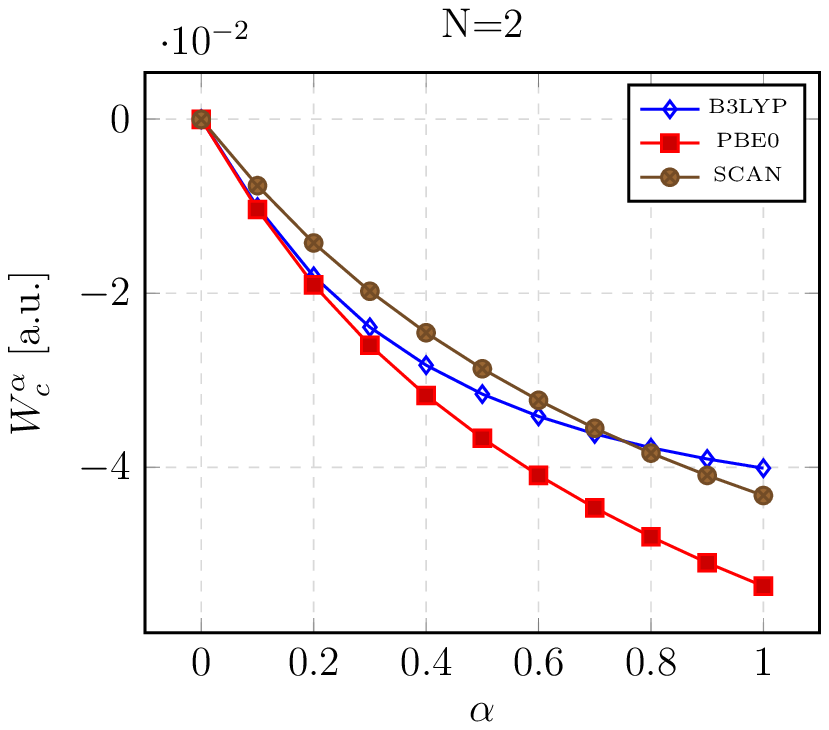} \hspace{0.5cm}%
 	\includegraphics[scale=0.55]{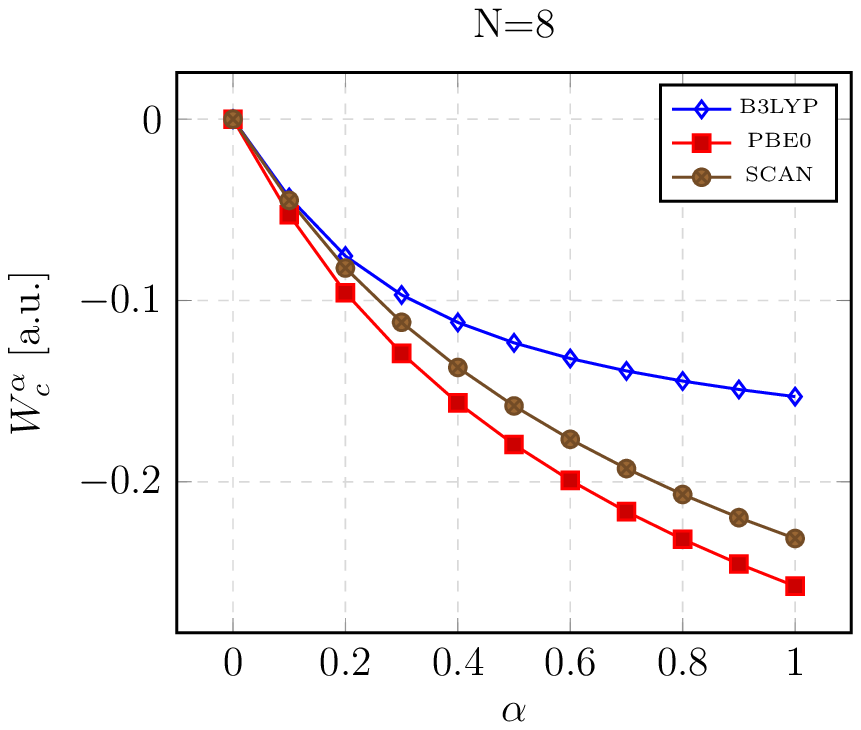}\hspace{0.5cm}
 	\includegraphics[scale=0.55]{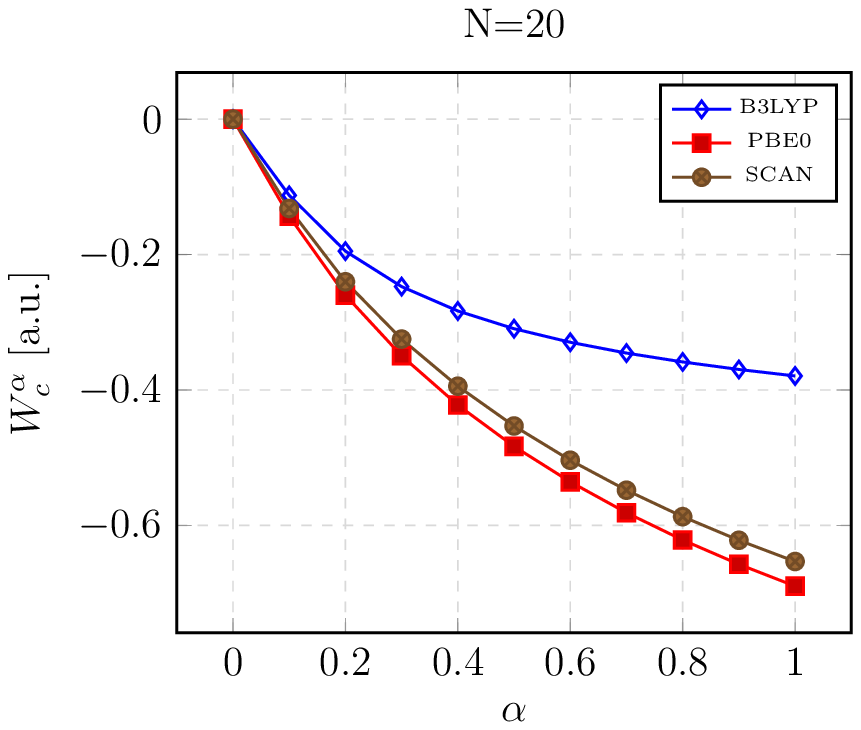} \vspace{0.5cm}\\
 	\includegraphics[scale=0.55]{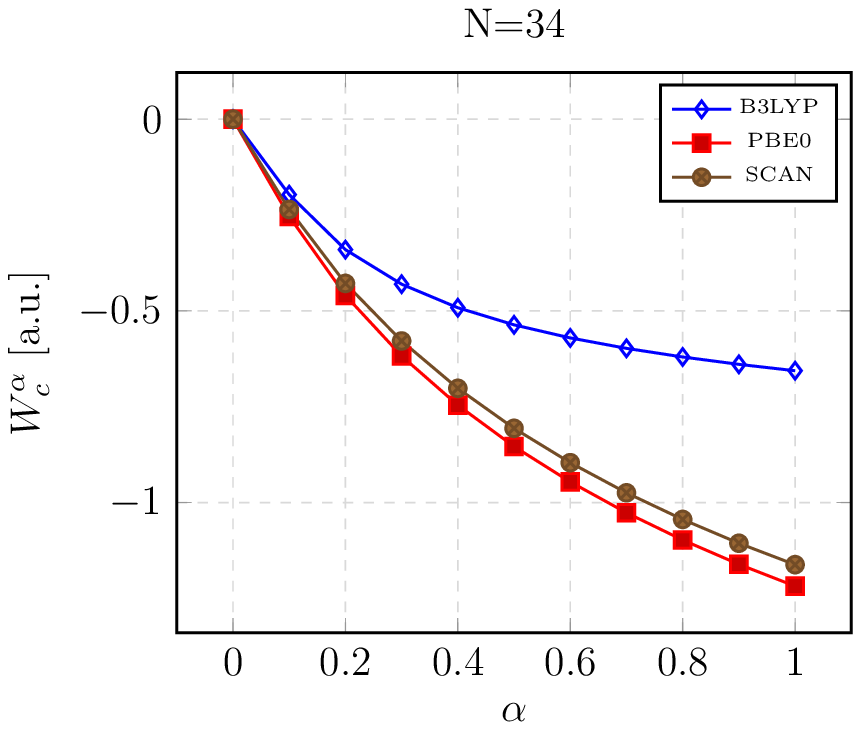} \hspace{0.5cm}%
 	\includegraphics[scale=0.55]{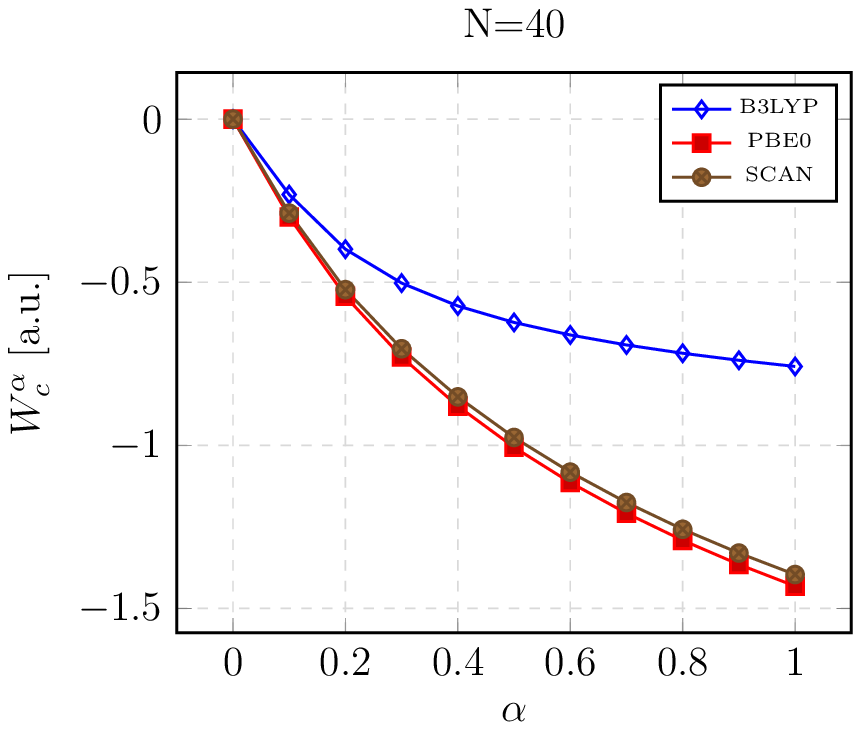}\hspace{0.5cm}
 	\includegraphics[scale=0.55]{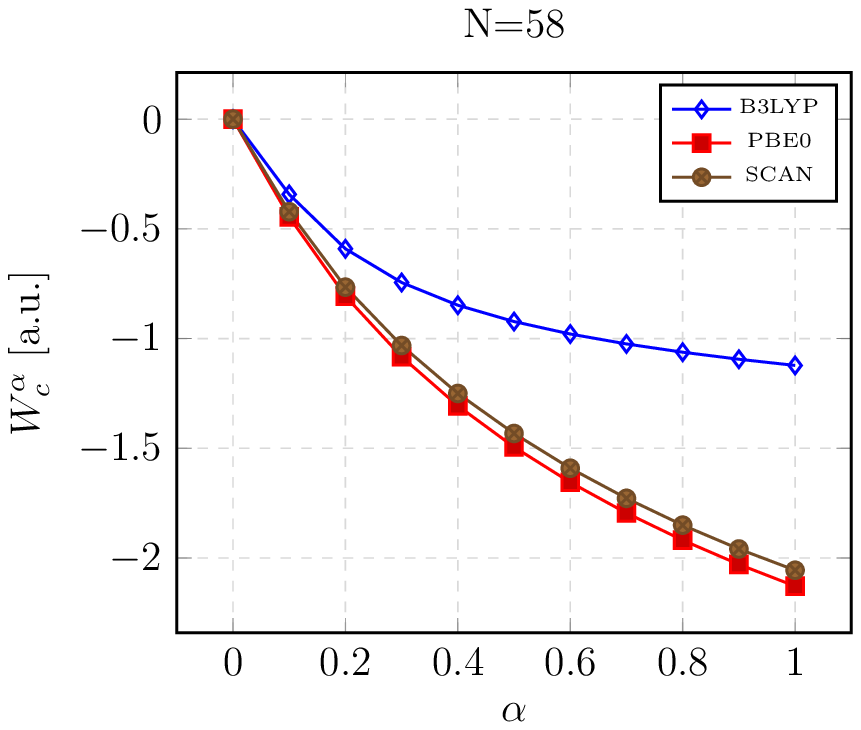}
 \end{figure*}
\subsection{Results for Hartree-Fock densities}
In this section we calculate $W_c^{\alpha}$ curves and $T_c$ for systems with two or more electrons using their Hartree-Fock (HF) densities. We first present results for two-electron systems and show that these are similar to those obtained in the previous section from near-exact-wavefunctions. This gives us confidence to use HF densities for other systems too and to draw conclusions on the basis of that. In the following we presents results for atoms and jellium clusters. The former have external potential proportional to $-\frac{1}{r}$ all over the space while the latter have external potential similar to the Hookium atom in their inner region and $-\frac{1}{r}$ in the outer region.
\subsubsection{Atoms}
In Table \ref{ex_ec_exd}, we have also displayed the exchange and correlation energies of H$^-$, He and the Hookium atom as obtained from their Hartree-Fock densities. It is observed that all the values obtained from the HF densities are nearly the same as those from the correlated wavefunctions. The same trend is observed for the values of $T_c$ obtained from the HF densities which are displayed in Table \ref{tc_exd} with the exception of H$^-$. We can therefore use HF densities for other neutral atoms studied in this paper and draw conclusion on the basis of that.
\par  Shown in Table \ref{ex_ec_hf_in} are the exchange and correlation energies for some spherical atoms obtained from their Hartree-Fock densities reproducing the results obtained earlier\cite{Scan}. In our calculations we have used densities that are given by self-consistent numerical calculation\cite{hf86}. The energies thus obtained are compared with their exact values.
The trends observed in the values of exchange and correlation energies are similar to those in Table \ref{ex_ec_exd}. The exchange energies given by the three functionals are nearly the same while there is a large variation in correlation energies. While the B3LYP functional overestimates the magnitude of correlation energies, the PBE0 and SCAN functionals underestimate it and give values that are close to each other.
\par In Table \ref{tc_hf_in}, we display the kinetic energy difference $T_c$ between the exact and Khon-Sham kinetic energies of the atoms studied in this section. It is observed that for all the atoms $T_c$ calculated for the three functionals differ. The B3LYP functional gives the largest value of $T_c$ and SCAN gives the smallest value.
\par  Next we display in Fig. \ref{wxc_hf_in} the  $W_c^{\alpha}$ versus $\alpha$ curves for Be, Ne and Ar. The curves have been obtained using the Hartree-Fock density for these atoms. We see that the PBE0 and the SCAN curves are close to each other while the B3LYP curve is below both of these. This is consistent with the total value of correlation energy displayed in Table \ref{ex_ec_hf_in}.
\par We conclude this section by looking at the exchange-correlation energies of atoms. Combining the exchange and correlation results so far from the exact densities in Section A or from the Hartree-Fock densities in Section B, we calculate the exchange-correlation energy of the atoms studies so far and display them in Table \ref{exc_atoms}. This is done to demonstrate the cancellation of errors between the exchange and correlation energies calculated using density functionals. It is evident that the SCAN functional give the best such cancellation of errors.
\subsubsection{Jellium spheres}
Having analyzed the functionals for atomic systems, where the external potential is proportional to $-\frac{1}{r}$, we now extend out study to neutral jellium spheres. In these spheres N electrons are moving in the potential of a uniformly charged sphere of carrying positive charge N with its charge density given as $\frac{3}{4\pi r_s^3}$. The radius of the sphere is $N^{1/3} r_s$ and the external potential is given as\cite{Matt} 
$$
    V(r)= 
    \begin{cases}
    	-\frac{3}{2}\frac{N^{\frac{2}{3}}}{r_s} 
    	+\frac{1}{2}\frac{r^2}{r_s^3},& \text{if } r\leq N^{\frac{1}{3}} r_s\\
    	-\frac{N^{\frac{2}{3}}}{r},              & \text{otherwise}
    	\end{cases}
$$ 	
    The parameter $r_s$ is known as the Wigner-Seitz radius\cite{Ashc} for the metal whose clusters are represented by the jellium spheres. We have taken $r_s=4$ which is close to the value of $r_s$ for sodium metal. Thus these systems resemble the Hookium atom studied in the previous section.
    \par The density that we employ to calculate the exchange and correlation energies is obtained by solving the Kohn-Sham equation for these systems employing the Harbola-Sahni exchange potential\cite{Harsa, QDFT}. The resulting densities are known to be essentially the same as Hartree-Fock densities.
 \par Shown in Table \ref{ex_ec_hf_cl} are the exchange and correlation energies per electron for the magic clusters with N= 2, 8, 18, 20, 34, 40 and 58. These are spheres with full occupancy of orbitals and represents magic clusters \cite{Knight1}. It is evident from the Table that the exchange energies per electron given by the three functionals are almost the same and close to the exact ones. On the other hand the correlation energies per electron vary quite significantly among the three functionals with the SCAN energy lying between that given by the B3LYP and the PBE0 functionals. The difference in the correlation energies arising from the three functionals is similar to that for atoms but is more accentuated in the clusters. These energies are compared with the exact correlation energies for these systems calculated \cite{Sottile_PRB.64.045105} using Monte-Carlo methods. We observe that the SCAN functional gives results closest to the exact values of energy per electron. As the sphere size grows, the PBE0 and SCAN values become close. The energy results farthest from the exact values are those for the B3LYP functional. This is expected since LYP functional
 is based on Colle-Salvetti\cite{Colle1975} ansatz that does not satisfy certain exact conditions\cite{Ranbir,Tao63} and therefore does not give\cite{Tao63} correlation energy per electron correctly in the limit of homogeneous electron gas.
\par To show cancellation of errors, as done for atoms in Table \ref{exc_atoms}, we show in Table \ref{exc_jc} the sum of exchange and correlation energies per electron for jellium spheres. Again we find that the SCAN functional gives best error cancellation between exchange and correlation energies of these systems.	
\par Next in Table \ref{tc_hf_cl}, we show $T_c$ as calculated from Eq. \ref{kico} for these clusters. Again trends are similar to those for the atoms but on a larger scale. However, PBE0 and SCAN functionals give results that are close to each other. We note that $T_c$ per electron for the bulk is 0.0082 a.u. (This value is obtained by using the virial relation\cite{Wang}) and all three functionals differ from it significantly.
\par In Fig. \ref{wxc_hf_cl} we show the $W_c^{\alpha}$ curves with respect to $\alpha$ for densities of spheres with N=2, 8, 20, 34, 40 and 58. It is seen that the curves for the PBE0 and SCAN functionals are very close and differ significantly from the B3LYP curve. For the N=2 sphere, the curves show a trend similar to that for the Hookium atom. 
\section{Summary}
In this paper, we have calculated the exchange and correlation energies for spherical systems with different forms of the external potential by employing the B3LYP, PBE0 and SCAN functionals and compared the results. To get further insight into the functionals, we have also plotted the $W_c^{\alpha}$ curves as a function of $\alpha$ for the three functionals and extracted from it $T_c$ $-$ the difference between the exact and Kohn-Sham kinetic energy $-$ for these functionals. Our study indicates that these functionals are accurate and give similar results for the exchange energies but can differ significantly in their values for the correlation energies and $T_c$. The differences are large for jellium spheres where the electron density is more spread out. We believe that our study nicely brings out the difference among the three functionals and will aid in analysis and development of newer functionals.  
\section*{Acknowledgement}
Communication and exchange of data with Prof. J. P. Perdew is gratefully acknowledged. Discussions with Ashish Kumar are also acknowledged. We also thank one of the anonymous referee for bringing to our attention reference that gives the exact correlation energies for jellium spheres\cite{Sottile_PRB.64.045105}.
\appendix
\section*{Appendix}
In this appendix we give details of how Eq. (\ref{wxca}) of the text is obtained using scaling relations. The derivation is essentially a detailed reproduction of the scaling arguments of Levy and Perdew\cite{LP85}.
\par
 The exchange-correlation energy for $\alpha V_{ee}$ interaction is calculated through the Hellman-Feynman theorem as 
\begin{equation}
E_{xc}^{\alpha}[\rho]= \int_0^{\alpha} W_{xc}^{\beta} d\beta   \label{excaa}
\end{equation}
with $W_{xc}^{\alpha}=\langle \Psi^{\alpha} |V_{ee}| \Psi^{\alpha} \rangle-\frac{1}{2} \iint \frac{\rho(\vec{r})\rho(\vec{r}')}{|\vec{r}-\vec{r}'|} d\vec{r}d\vec{r}'$, where $\Psi^{\alpha}$ is the solution of the Schr$\ddot{\text{o}}$dinger equation 
\begin{equation}
	[\hat{T} + \alpha V_{ee} + V_{ext}^{\alpha}] \Psi^{\alpha}(\vec{r})=E^{\alpha} \Psi^{\alpha}(\vec{r}) \label{sceqa1}
\end{equation}    
and gives the density $\rho(\vec{r})$. Notice that $V_{ext}^{\alpha}$ depends on $\alpha$ such that the density given by $\Psi^{\alpha}$ remains the same irrespective of the value of $\alpha$. Now take $\vec{r}=\lambda \vec{z}$ to write Eq. (\ref{sceqa1}) as
 \begin{equation*}
 \Big[\frac{\hat{T}(\vec{z})}{\lambda^2} + \frac{\alpha}{\lambda}V_{ee}(\vec{z}) + V_{ext}^{\alpha}(\lambda \vec{z})\Big]\Psi^{\alpha}(\lambda \vec{z}) = E^{\alpha} \Psi^{\alpha}(\lambda \vec{z})
 \end{equation*}
 or
 \begin{equation}
 \Big[\hat{T}(\vec{z}) + \lambda \alpha V_{ee}(\vec{z})+\lambda^2 V_{ext}^{\alpha}(\lambda \vec{z})\Big]\Psi^{\alpha}(\lambda \vec{z})=\lambda^2 E^{\alpha} \Psi^{\alpha}(\lambda \vec{z}). \label{sceqa2}
 \end{equation}
 Taking $\lambda = \frac{1}{\alpha}$, gives
 \begin{equation}
\Big [\hat{T}(\vec{z}) + V_{ee}(\vec{z})+\frac{1}{\alpha^2} V_{ext}^{\alpha}\Big(\frac{\vec{z}}{\alpha}\Big)\Big]\Psi^{\alpha}\Big(\frac{ \vec{z}}{\alpha}\big)=\frac{1}{\alpha^2} E^{\alpha} \Psi^{\alpha}\Big(\frac{\vec{z}}{\alpha}\Big). \label{sceqa3}
 \end{equation}
 Properly normalized $\Psi^{\alpha}(\frac{\vec{z}}{\alpha})$ is given as $(\frac{1}{\alpha})^{\frac{3N}{2}} \Psi^{\alpha}(\frac{\vec{z}}{\alpha})$. Thus it leads to the density $(\frac{1}{\alpha})^3 \rho(\frac{\vec{z}}{\alpha})$ which will be denoted as $\rho_{\frac{1}{\alpha}}(\vec{z})$. Evidently, $(\frac{1}{\alpha})^{\frac{3N}{2}} \Psi^{\alpha}(\frac{\vec{z}}{\alpha})$ is that wavefunction that gives the density $\rho_{\frac{1}{\alpha}}$ and minimizes $\langle \hat{T} + V_{ee} \rangle$. This then gives the exchange-correlation energy for the corresponding density $\rho_{\frac{1}{\alpha}}(\vec{r})=\frac{1}{\alpha^3}\rho(\frac{\vec{r}}{\alpha})$ as
 \begin{equation}
 \begin{split}
 E_{xc}[\rho_{\frac{1}{\alpha}}]&= \frac{1}{\alpha^{3N}} \langle \Psi^{\alpha}\Big(\frac{\vec{r}}{\alpha}\Big)|\hat{T} + V_{ee}|\Psi^{\alpha}\Big(\frac{\vec{r}}{\alpha}\Big)\rangle \\ 
 &- \frac{1}{2} \iint \frac{\rho_{\frac{1}{\alpha}}(\vec{r}) \rho_{\frac{1}{\alpha}(\vec{r}')}}{|\vec{r}-\vec{r}'|} d\vec{r} d\vec{r}' \\ 
 &- \frac{1}{\alpha^{3N}} \langle \Psi^{0}\Big(\frac{\vec{r}}{\alpha}\Big)|\hat{T}|\Psi^{0}\Big(\frac{\vec{r}}{\alpha}\Big)\rangle 
\end{split}
\label{excra}
 \end{equation}
 where $\Psi^0(\vec{r})$ is the Kohn-Sham wavefunction (a Slater determinant) i.e. the solution of Eq. (\ref{sceqa1}) with $\alpha=0$. Keep in mind that in calculating the expectation values in Eq. (\ref{excra}), the integrations are performed over $\vec{r}$ and $\vec{r}'$ variables. Changing variables to $\vec{z}=\frac{\vec{r}}{\alpha}$, we get
 \begin{eqnarray}
 E_{xc}[\rho_{\frac{1}{\alpha}}] &=& \langle \Psi^{\alpha}(\vec{z})|\frac{\hat{T}(\vec{z})}{\alpha^2} + \frac{V_{ee}(\vec{z})}{\alpha}|\Psi^{\alpha}(\vec{z}) \rangle \\ \nonumber
 &-& \frac{1}{2}\frac{1}{\alpha} \iint \frac{\rho(\vec{z})\rho(\vec{z}')}{|\vec{z}-\vec{a}'|} d\vec{z} d\vec{z}' \\ \nonumber
 &-& \frac{1}{\alpha^2} \langle \Psi^0(\vec{z}) |T(\vec{z})| \Psi^0(\vec{z}) \rangle \label{excra1}
 \end{eqnarray} 
 or 
 \begin{eqnarray}
 \alpha^2 E_{xc}[\rho_{\frac{1}{\alpha}}]= \langle \Psi^{\alpha}|\hat{T} &+&\alpha V_{ee}| \Psi^{\alpha} \rangle \\ \nonumber
 &-& \alpha E_{H}[\rho] - T_s[\rho] \label{excra2}
 \end{eqnarray}
 From the definition of exchange-correlation energy, the right
 hand side of Eq. (\ref{excra1}) is the exchange-correlation energy $E_{xc}^{\alpha}[\rho]$ for the system described by the Hamiltonian of Eq. (\ref{scal}). Furthermore, from Eq. (\ref{excaa}) we get
 $\frac{d}{d\alpha}(E_{xc}^{\alpha}[\rho])=W_{xc}^{\alpha}$, so
 $W_{xc}^{\alpha} = \frac{d}{d\alpha}(\alpha^2 E_{xc}^{\alpha}[\rho_{\frac{1}{\alpha}}])$ using Eq. (\ref{excra1}).

%\begin{acknowledgments}
%\end{acknowledgments}

%\bibliography{shorttitles,citations}

\begin{thebibliography}{65}%
\makeatletter
\providecommand \@ifxundefined [1]{%
 \@ifx{#1\undefined}
}%
\providecommand \@ifnum [1]{%
 \ifnum #1\expandafter \@firstoftwo
 \else \expandafter \@secondoftwo
 \fi
}%
\providecommand \@ifx [1]{%
 \ifx #1\expandafter \@firstoftwo
 \else \expandafter \@secondoftwo
 \fi
}%
\providecommand \natexlab [1]{#1}%
\providecommand \enquote  [1]{``#1''}%
\providecommand \bibnamefont  [1]{#1}%
\providecommand \bibfnamefont [1]{#1}%
\providecommand \citenamefont [1]{#1}%
\providecommand \href@noop [0]{\@secondoftwo}%
\providecommand \href [0]{\begingroup \@sanitize@url \@href}%
\providecommand \@href[1]{\@@startlink{#1}\@@href}%
\providecommand \@@href[1]{\endgroup#1\@@endlink}%
\providecommand \@sanitize@url [0]{\catcode `\\12\catcode `\$12\catcode
  `\&12\catcode `\#12\catcode `\^12\catcode `\_12\catcode `\%12\relax}%
\providecommand \@@startlink[1]{}%
\providecommand \@@endlink[0]{}%
\providecommand \url  [0]{\begingroup\@sanitize@url \@url }%
\providecommand \@url [1]{\endgroup\@href {#1}{\urlprefix }}%
\providecommand \urlprefix  [0]{URL }%
\providecommand \Eprint [0]{\href }%
\providecommand \doibase [0]{http://dx.doi.org/}%
\providecommand \selectlanguage [0]{\@gobble}%
\providecommand \bibinfo  [0]{\@secondoftwo}%
\providecommand \bibfield  [0]{\@secondoftwo}%
\providecommand \translation [1]{[#1]}%
\providecommand \BibitemOpen [0]{}%
\providecommand \bibitemStop [0]{}%
\providecommand \bibitemNoStop [0]{.\EOS\space}%
\providecommand \EOS [0]{\spacefactor3000\relax}%
\providecommand \BibitemShut  [1]{\csname bibitem#1\endcsname}%
\let\auto@bib@innerbib\@empty
%</preamble>
\bibitem [{\citenamefont {Hohenberg}\ and\ \citenamefont {Kohn}(1964)}]{Hohn}%
  \BibitemOpen
  \bibfield  {author} {\bibinfo {author} {\bibfnamefont {P.}~\bibnamefont
  {Hohenberg}}\ and\ \bibinfo {author} {\bibfnamefont {W.}~\bibnamefont
  {Kohn}},\ }\href {http://journals.aps.org/pr/pdf/10.1103/PhysRev.136.B864}
  {\bibfield  {journal} {\bibinfo  {journal} {Phys. Rev.}\ }\textbf {\bibinfo
  {volume} {136}},\ \bibinfo {pages} {B864} (\bibinfo {year}
  {1964})}\BibitemShut {NoStop}%
\bibitem [{\citenamefont {Parr}\ and\ \citenamefont {Yang}(1995)}]{Yang}%
  \BibitemOpen
  \bibfield  {author} {\bibinfo {author} {\bibfnamefont {R.~G.}\ \bibnamefont
  {Parr}}\ and\ \bibinfo {author} {\bibfnamefont {W.}~\bibnamefont {Yang}},\
  }\href
  {https://global.oup.com/academic/product/density-functional-theory-of-atoms-and-molecules-9780195092769?cc=in&lang=en&}
  {\emph {\bibinfo {title} {Density-Functional Theory of Atoms and
  Molecules}}}\ (\bibinfo  {publisher} {Oxford Science Publications},\ \bibinfo
  {year} {1995})\BibitemShut {NoStop}%
\bibitem [{\citenamefont {Tong}\ and\ \citenamefont {Sham}(1966)}]{Tong}%
  \BibitemOpen
  \bibfield  {author} {\bibinfo {author} {\bibfnamefont {B.~Y.}\ \bibnamefont
  {Tong}}\ and\ \bibinfo {author} {\bibfnamefont {L.~J.}\ \bibnamefont
  {Sham}},\ }\href {\doibase 10.1103/PhysRev.144.1} {\bibfield  {journal}
  {\bibinfo  {journal} {Phys. Rev.}\ }\textbf {\bibinfo {volume} {144}},\
  \bibinfo {pages} {1} (\bibinfo {year} {1966})}\BibitemShut {NoStop}%
\bibitem [{\citenamefont {Kohn}\ and\ \citenamefont {Sham}(1965)}]{Kohn}%
  \BibitemOpen
  \bibfield  {author} {\bibinfo {author} {\bibfnamefont {W.}~\bibnamefont
  {Kohn}}\ and\ \bibinfo {author} {\bibfnamefont {L.~J.}\ \bibnamefont
  {Sham}},\ }\href {http://journals.aps.org/pr/pdf/10.1103/PhysRev.140.A1133}
  {\bibfield  {journal} {\bibinfo  {journal} {Phys. Rev.}\ }\textbf {\bibinfo
  {volume} {140}},\ \bibinfo {pages} {A1133} (\bibinfo {year}
  {1965})}\BibitemShut {NoStop}%
\bibitem [{\citenamefont {Gunnarsson}\ and\ \citenamefont
  {Lundqvist}(1976)}]{Gun1}%
  \BibitemOpen
  \bibfield  {author} {\bibinfo {author} {\bibfnamefont {O.}~\bibnamefont
  {Gunnarsson}}\ and\ \bibinfo {author} {\bibfnamefont {B.~I.}\ \bibnamefont
  {Lundqvist}},\ }\href
  {http://journals.aps.org/prb/pdf/10.1103/PhysRevB.13.4274} {\bibfield
  {journal} {\bibinfo  {journal} {Phys. Rev. B}\ }\textbf {\bibinfo {volume}
  {13}},\ \bibinfo {pages} {4274} (\bibinfo {year} {1976})}\BibitemShut
  {NoStop}%
\bibitem [{\citenamefont {Herman}\ \emph {et~al.}(1969)\citenamefont {Herman},
  \citenamefont {Van~Dyke},\ and\ \citenamefont {Ortenburger}}]{Herman}%
  \BibitemOpen
  \bibfield  {author} {\bibinfo {author} {\bibfnamefont {F.}~\bibnamefont
  {Herman}}, \bibinfo {author} {\bibfnamefont {J.~P.}\ \bibnamefont
  {Van~Dyke}}, \ and\ \bibinfo {author} {\bibfnamefont {I.~B.}\ \bibnamefont
  {Ortenburger}},\ }\href {\doibase 10.1103/PhysRevLett.22.807} {\bibfield
  {journal} {\bibinfo  {journal} {Phys. Rev. Lett.}\ }\textbf {\bibinfo
  {volume} {22}},\ \bibinfo {pages} {807} (\bibinfo {year} {1969})}\BibitemShut
  {NoStop}%
\bibitem [{\citenamefont {Sham}(1971)}]{Sham1971}%
  \BibitemOpen
  \bibfield  {author} {\bibinfo {author} {\bibfnamefont {L.~J.}\ \bibnamefont
  {Sham}},\ }\enquote {\bibinfo {title} {Approximations of the exchange and
  correlation potentials},}\ in\ \href {\doibase 10.1007/978-1-4684-1890-3_36}
  {\emph {\bibinfo {booktitle} {Computational Methods in Band Theory:
  Proceedings of a Conference held at the IBM Thomas J. Watson Research Center,
  Yorktown Heights, New York, May 14--15, 1970, under the joint sponsorship of
  IBM and the American Physical Society}}},\ \bibinfo {editor} {edited by\
  \bibinfo {editor} {\bibfnamefont {P.~M.}\ \bibnamefont {Marcus}}, \bibinfo
  {editor} {\bibfnamefont {J.~F.}\ \bibnamefont {Janak}}, \ and\ \bibinfo
  {editor} {\bibfnamefont {A.~R.}\ \bibnamefont {Williams}}}\ (\bibinfo
  {publisher} {Springer US},\ \bibinfo {address} {Boston, MA},\ \bibinfo {year}
  {1971})\ pp.\ \bibinfo {pages} {458--468}\BibitemShut {NoStop}%
\bibitem [{\citenamefont {Gross}\ and\ \citenamefont {Dreizler}(1981)}]{Gross}%
  \BibitemOpen
  \bibfield  {author} {\bibinfo {author} {\bibfnamefont {E.~K.~U.}\
  \bibnamefont {Gross}}\ and\ \bibinfo {author} {\bibfnamefont {R.~M.}\
  \bibnamefont {Dreizler}},\ }\href {\doibase 10.1007/BF01413038} {\bibfield
  {journal} {\bibinfo  {journal} {Z. Phys. A: At. Nucl.}\ }\textbf {\bibinfo
  {volume} {302}},\ \bibinfo {pages} {103} (\bibinfo {year}
  {1981})}\BibitemShut {NoStop}%
\bibitem [{\citenamefont {Perdew}(1985)}]{Perd85}%
  \BibitemOpen
  \bibfield  {author} {\bibinfo {author} {\bibfnamefont {J.~P.}\ \bibnamefont
  {Perdew}},\ }\href {\doibase 10.1103/PhysRevLett.55.1665} {\bibfield
  {journal} {\bibinfo  {journal} {Phys. Rev. Lett.}\ }\textbf {\bibinfo
  {volume} {55}},\ \bibinfo {pages} {1665} (\bibinfo {year}
  {1985})}\BibitemShut {NoStop}%
\bibitem [{\citenamefont {Becke}(1988)}]{Bec88}%
  \BibitemOpen
  \bibfield  {author} {\bibinfo {author} {\bibfnamefont {A.~D.}\ \bibnamefont
  {Becke}},\ }\href {\doibase 10.1103/PhysRevA.38.3098} {\bibfield  {journal}
  {\bibinfo  {journal} {Phys. Rev. A}\ }\textbf {\bibinfo {volume} {38}},\
  \bibinfo {pages} {3098} (\bibinfo {year} {1988})}\BibitemShut {NoStop}%
\bibitem [{\citenamefont {Wigner}\ and\ \citenamefont {Seitz}(1933)}]{Wigner1}%
  \BibitemOpen
  \bibfield  {author} {\bibinfo {author} {\bibfnamefont {E.}~\bibnamefont
  {Wigner}}\ and\ \bibinfo {author} {\bibfnamefont {F.}~\bibnamefont {Seitz}},\
  }\href {\doibase 10.1103/PhysRev.43.804} {\bibfield  {journal} {\bibinfo
  {journal} {Phys. Rev.}\ }\textbf {\bibinfo {volume} {43}},\ \bibinfo {pages}
  {804} (\bibinfo {year} {1933})}\BibitemShut {NoStop}%
\bibitem [{\citenamefont {Wigner}\ and\ \citenamefont {Seitz}(1934)}]{Wigner2}%
  \BibitemOpen
  \bibfield  {author} {\bibinfo {author} {\bibfnamefont {E.}~\bibnamefont
  {Wigner}}\ and\ \bibinfo {author} {\bibfnamefont {F.}~\bibnamefont {Seitz}},\
  }\href {\doibase 10.1103/PhysRev.46.509} {\bibfield  {journal} {\bibinfo
  {journal} {Phys. Rev.}\ }\textbf {\bibinfo {volume} {46}},\ \bibinfo {pages}
  {509} (\bibinfo {year} {1934})}\BibitemShut {NoStop}%
\bibitem [{\citenamefont {Wigner}(1934)}]{Wigner3}%
  \BibitemOpen
  \bibfield  {author} {\bibinfo {author} {\bibfnamefont {E.}~\bibnamefont
  {Wigner}},\ }\href {\doibase 10.1103/PhysRev.46.1002} {\bibfield  {journal}
  {\bibinfo  {journal} {Phys. Rev.}\ }\textbf {\bibinfo {volume} {46}},\
  \bibinfo {pages} {1002} (\bibinfo {year} {1934})}\BibitemShut {NoStop}%
\bibitem [{\citenamefont {Gell-Mann}\ and\ \citenamefont
  {Brueckner}(1957)}]{Gelbru}%
  \BibitemOpen
  \bibfield  {author} {\bibinfo {author} {\bibfnamefont {M.}~\bibnamefont
  {Gell-Mann}}\ and\ \bibinfo {author} {\bibfnamefont {K.~A.}\ \bibnamefont
  {Brueckner}},\ }\href {\doibase 10.1103/PhysRev.106.364} {\bibfield
  {journal} {\bibinfo  {journal} {Phys. Rev.}\ }\textbf {\bibinfo {volume}
  {106}},\ \bibinfo {pages} {364} (\bibinfo {year} {1957})}\BibitemShut
  {NoStop}%
\bibitem [{\citenamefont {Ma}\ and\ \citenamefont {Brueckner}(1968)}]{Mabru}%
  \BibitemOpen
  \bibfield  {author} {\bibinfo {author} {\bibfnamefont {S.-K.}\ \bibnamefont
  {Ma}}\ and\ \bibinfo {author} {\bibfnamefont {K.~A.}\ \bibnamefont
  {Brueckner}},\ }\href {\doibase 10.1103/PhysRev.165.18} {\bibfield  {journal}
  {\bibinfo  {journal} {Phys. Rev.}\ }\textbf {\bibinfo {volume} {165}},\
  \bibinfo {pages} {18} (\bibinfo {year} {1968})}\BibitemShut {NoStop}%
\bibitem [{\citenamefont {Ceperley}\ and\ \citenamefont {Alder}(1980)}]{Cepal}%
  \BibitemOpen
  \bibfield  {author} {\bibinfo {author} {\bibfnamefont {D.~M.}\ \bibnamefont
  {Ceperley}}\ and\ \bibinfo {author} {\bibfnamefont {B.~J.}\ \bibnamefont
  {Alder}},\ }\href {\doibase 10.1103/PhysRevLett.45.566} {\bibfield  {journal}
  {\bibinfo  {journal} {Phys. Rev. Lett.}\ }\textbf {\bibinfo {volume} {45}},\
  \bibinfo {pages} {566} (\bibinfo {year} {1980})}\BibitemShut {NoStop}%
\bibitem [{\citenamefont {Langreth}\ and\ \citenamefont {Mehl}(1983)}]{Lanme}%
  \BibitemOpen
  \bibfield  {author} {\bibinfo {author} {\bibfnamefont {D.~C.}\ \bibnamefont
  {Langreth}}\ and\ \bibinfo {author} {\bibfnamefont {M.~J.}\ \bibnamefont
  {Mehl}},\ }\href {\doibase 10.1103/PhysRevB.28.1809} {\bibfield  {journal}
  {\bibinfo  {journal} {Phys. Rev. B}\ }\textbf {\bibinfo {volume} {28}},\
  \bibinfo {pages} {1809} (\bibinfo {year} {1983})}\BibitemShut {NoStop}%
\bibitem [{\citenamefont {Langreth}\ and\ \citenamefont
  {Perdew}(1980)}]{Lanpr}%
  \BibitemOpen
  \bibfield  {author} {\bibinfo {author} {\bibfnamefont {D.~C.}\ \bibnamefont
  {Langreth}}\ and\ \bibinfo {author} {\bibfnamefont {J.~P.}\ \bibnamefont
  {Perdew}},\ }\href {\doibase 10.1103/PhysRevB.21.5469} {\bibfield  {journal}
  {\bibinfo  {journal} {Phys. Rev. B}\ }\textbf {\bibinfo {volume} {21}},\
  \bibinfo {pages} {5469} (\bibinfo {year} {1980})}\BibitemShut {NoStop}%
\bibitem [{\citenamefont {Perdew}\ and\ \citenamefont {Yue}(1986)}]{Perwa}%
  \BibitemOpen
  \bibfield  {author} {\bibinfo {author} {\bibfnamefont {J.~P.}\ \bibnamefont
  {Perdew}}\ and\ \bibinfo {author} {\bibfnamefont {W.}~\bibnamefont {Yue}},\
  }\href {\doibase 10.1103/PhysRevB.33.8800} {\bibfield  {journal} {\bibinfo
  {journal} {Phys. Rev. B}\ }\textbf {\bibinfo {volume} {33}},\ \bibinfo
  {pages} {8800} (\bibinfo {year} {1986})}\BibitemShut {NoStop}%
\bibitem [{\citenamefont {Perdew}\ \emph
  {et~al.}(1996{\natexlab{a}})\citenamefont {Perdew}, \citenamefont {Burke},\
  and\ \citenamefont {Ernzerhof}}]{GGA}%
  \BibitemOpen
  \bibfield  {author} {\bibinfo {author} {\bibfnamefont {J.~P.}\ \bibnamefont
  {Perdew}}, \bibinfo {author} {\bibfnamefont {K.}~\bibnamefont {Burke}}, \
  and\ \bibinfo {author} {\bibfnamefont {M.}~\bibnamefont {Ernzerhof}},\ }\href
  {http://journals.aps.org/prl/pdf/10.1103/PhysRevLett.77.3865} {\bibfield
  {journal} {\bibinfo  {journal} {Phys. Rev. Lett.}\ }\textbf {\bibinfo
  {volume} {77}},\ \bibinfo {pages} {3865} (\bibinfo {year}
  {1996}{\natexlab{a}})}\BibitemShut {NoStop}%
\bibitem [{\citenamefont {Svendsen}\ and\ \citenamefont {von
  Barth}(1995)}]{Sven}%
  \BibitemOpen
  \bibfield  {author} {\bibinfo {author} {\bibfnamefont {P.~S.}\ \bibnamefont
  {Svendsen}}\ and\ \bibinfo {author} {\bibfnamefont {U.}~\bibnamefont {von
  Barth}},\ }\href {\doibase 10.1002/qua.560560421} {\bibfield  {journal}
  {\bibinfo  {journal} {Int. J. Quantum Chem.}\ }\textbf {\bibinfo {volume}
  {56}},\ \bibinfo {pages} {351} (\bibinfo {year} {1995})}\BibitemShut
  {NoStop}%
\bibitem [{\citenamefont {Svendsen}\ and\ \citenamefont {von
  Barth}(1996)}]{PSBarth}%
  \BibitemOpen
  \bibfield  {author} {\bibinfo {author} {\bibfnamefont {P.~S.}\ \bibnamefont
  {Svendsen}}\ and\ \bibinfo {author} {\bibfnamefont {U.}~\bibnamefont {von
  Barth}},\ }\href {\doibase 10.1103/PhysRevB.54.17402} {\bibfield  {journal}
  {\bibinfo  {journal} {Phys. Rev. B}\ }\textbf {\bibinfo {volume} {54}},\
  \bibinfo {pages} {17402} (\bibinfo {year} {1996})}\BibitemShut {NoStop}%
\bibitem [{\citenamefont {Becke}(1993{\natexlab{a}})}]{Beck}%
  \BibitemOpen
  \bibfield  {author} {\bibinfo {author} {\bibfnamefont {A.~D.}\ \bibnamefont
  {Becke}},\ }\href {http://aip.scitation.org/doi/pdf/10.1063/1.464304}
  {\bibfield  {journal} {\bibinfo  {journal} {J. Chem. Phys.}\ }\textbf
  {\bibinfo {volume} {98}},\ \bibinfo {pages} {1372} (\bibinfo {year}
  {1993}{\natexlab{a}})}\BibitemShut {NoStop}%
\bibitem [{\citenamefont {Becke}(1993{\natexlab{b}})}]{Becke}%
  \BibitemOpen
  \bibfield  {author} {\bibinfo {author} {\bibfnamefont {A.~D.}\ \bibnamefont
  {Becke}},\ }\href {http://aip.scitation.org/doi/pdf/10.1063/1.464913}
  {\bibfield  {journal} {\bibinfo  {journal} {J. Chem. Phys.}\ }\textbf
  {\bibinfo {volume} {98}},\ \bibinfo {pages} {5648} (\bibinfo {year}
  {1993}{\natexlab{b}})}\BibitemShut {NoStop}%
\bibitem [{\citenamefont {Lee}\ \emph {et~al.}(1988)\citenamefont {Lee},
  \citenamefont {Yang},\ and\ \citenamefont {Parr}}]{LYP}%
  \BibitemOpen
  \bibfield  {author} {\bibinfo {author} {\bibfnamefont {C.}~\bibnamefont
  {Lee}}, \bibinfo {author} {\bibfnamefont {W.}~\bibnamefont {Yang}}, \ and\
  \bibinfo {author} {\bibfnamefont {R.~G.}\ \bibnamefont {Parr}},\ }\href
  {http://journals.aps.org/prb/pdf/10.1103/PhysRevB.37.785} {\bibfield
  {journal} {\bibinfo  {journal} {Phys. Rev. B}\ }\textbf {\bibinfo {volume}
  {37}},\ \bibinfo {pages} {785} (\bibinfo {year} {1988})}\BibitemShut
  {NoStop}%
\bibitem [{\citenamefont {Perdew}\ \emph
  {et~al.}(1996{\natexlab{b}})\citenamefont {Perdew}, \citenamefont
  {Ernzerhof},\ and\ \citenamefont {Burke}}]{Perd}%
  \BibitemOpen
  \bibfield  {author} {\bibinfo {author} {\bibfnamefont {J.~P.}\ \bibnamefont
  {Perdew}}, \bibinfo {author} {\bibfnamefont {M.}~\bibnamefont {Ernzerhof}}, \
  and\ \bibinfo {author} {\bibfnamefont {K.}~\bibnamefont {Burke}},\ }\href
  {http://aip.scitation.org/doi/pdf/10.1063/1.472933} {\bibfield  {journal}
  {\bibinfo  {journal} {J. Chem. Phys.}\ }\textbf {\bibinfo {volume} {105}},\
  \bibinfo {pages} {9982} (\bibinfo {year} {1996}{\natexlab{b}})}\BibitemShut
  {NoStop}%
\bibitem [{\citenamefont {Adamo}\ and\ \citenamefont {Barone}(1999)}]{PBE0}%
  \BibitemOpen
  \bibfield  {author} {\bibinfo {author} {\bibfnamefont {C.}~\bibnamefont
  {Adamo}}\ and\ \bibinfo {author} {\bibfnamefont {V.}~\bibnamefont {Barone}},\
  }\href {\doibase 10.1063/1.478522} {\bibfield  {journal} {\bibinfo  {journal}
  {J. Chem. Phys.}\ }\textbf {\bibinfo {volume} {110}},\ \bibinfo {pages}
  {6158} (\bibinfo {year} {1999})}\BibitemShut {NoStop}%
\bibitem [{\citenamefont {Sun}\ \emph {et~al.}(2015)\citenamefont {Sun},
  \citenamefont {Ruzsinszky},\ and\ \citenamefont {Perdew}}]{Scan}%
  \BibitemOpen
  \bibfield  {author} {\bibinfo {author} {\bibfnamefont {J.}~\bibnamefont
  {Sun}}, \bibinfo {author} {\bibfnamefont {A.}~\bibnamefont {Ruzsinszky}}, \
  and\ \bibinfo {author} {\bibfnamefont {J.~P.}\ \bibnamefont {Perdew}},\
  }\href {http://journals.aps.org/prl/pdf/10.1103/PhysRevLett.115.036402}
  {\bibfield  {journal} {\bibinfo  {journal} {Phys. Rev. Lett.}\ }\textbf
  {\bibinfo {volume} {115}},\ \bibinfo {pages} {036402} (\bibinfo {year}
  {2015})}\BibitemShut {NoStop}%
\bibitem [{\citenamefont {Medvedev}\ \emph {et~al.}(2017)\citenamefont
  {Medvedev}, \citenamefont {Bushmarinov}, \citenamefont {Sun}, \citenamefont
  {Perdew},\ and\ \citenamefont {Lyssenko}}]{Medve}%
  \BibitemOpen
  \bibfield  {author} {\bibinfo {author} {\bibfnamefont {M.~G.}\ \bibnamefont
  {Medvedev}}, \bibinfo {author} {\bibfnamefont {I.~S.}\ \bibnamefont
  {Bushmarinov}}, \bibinfo {author} {\bibfnamefont {J.}~\bibnamefont {Sun}},
  \bibinfo {author} {\bibfnamefont {J.~P.}\ \bibnamefont {Perdew}}, \ and\
  \bibinfo {author} {\bibfnamefont {K.~A.}\ \bibnamefont {Lyssenko}},\ }\href
  {http://science.sciencemag.org/content/355/6320/49.full.pdf+html} {\bibfield
  {journal} {\bibinfo  {journal} {Science}\ }\textbf {\bibinfo {volume}
  {355}},\ \bibinfo {pages} {49} (\bibinfo {year} {2017})}\BibitemShut
  {NoStop}%
\bibitem [{\citenamefont {Peverati}\ and\ \citenamefont
  {Truhlar}(2014)}]{TruhF}%
  \BibitemOpen
  \bibfield  {author} {\bibinfo {author} {\bibfnamefont {R.}~\bibnamefont
  {Peverati}}\ and\ \bibinfo {author} {\bibfnamefont {D.~G.}\ \bibnamefont
  {Truhlar}},\ }\href
  {http://rsta.royalsocietypublishing.org/content/roypta/372/2011/20120476.full.pdf}
  {\bibfield  {journal} {\bibinfo  {journal} {Phil. Trans. R. Soc. A}\ }\textbf
  {\bibinfo {volume} {372}},\ \bibinfo {pages} {20120476} (\bibinfo {year}
  {2014})}\BibitemShut {NoStop}%
\bibitem [{\citenamefont {Knight}\ \emph {et~al.}(1984)\citenamefont {Knight},
  \citenamefont {Clemenger}, \citenamefont {de~Heer}, \citenamefont {Saunders},
  \citenamefont {Chou},\ and\ \citenamefont {Cohen}}]{Knight1}%
  \BibitemOpen
  \bibfield  {author} {\bibinfo {author} {\bibfnamefont {W.~D.}\ \bibnamefont
  {Knight}}, \bibinfo {author} {\bibfnamefont {K.}~\bibnamefont {Clemenger}},
  \bibinfo {author} {\bibfnamefont {W.~A.}\ \bibnamefont {de~Heer}}, \bibinfo
  {author} {\bibfnamefont {W.~A.}\ \bibnamefont {Saunders}}, \bibinfo {author}
  {\bibfnamefont {M.~Y.}\ \bibnamefont {Chou}}, \ and\ \bibinfo {author}
  {\bibfnamefont {M.~L.}\ \bibnamefont {Cohen}},\ }\href {\doibase
  10.1103/PhysRevLett.52.2141} {\bibfield  {journal} {\bibinfo  {journal}
  {Phys. Rev. Lett.}\ }\textbf {\bibinfo {volume} {52}},\ \bibinfo {pages}
  {2141} (\bibinfo {year} {1984})}\BibitemShut {NoStop}%
\bibitem [{\citenamefont {Brack}(1993)}]{Matt}%
  \BibitemOpen
  \bibfield  {author} {\bibinfo {author} {\bibfnamefont {M.}~\bibnamefont
  {Brack}},\ }\href {\doibase 10.1103/RevModPhys.65.677} {\bibfield  {journal}
  {\bibinfo  {journal} {Rev. Mod. Phys.}\ }\textbf {\bibinfo {volume} {65}},\
  \bibinfo {pages} {677} (\bibinfo {year} {1993})}\BibitemShut {NoStop}%
\bibitem [{\citenamefont {de~Heer}(1993)}]{Walt}%
  \BibitemOpen
  \bibfield  {author} {\bibinfo {author} {\bibfnamefont {W.~A.}\ \bibnamefont
  {de~Heer}},\ }\href {\doibase 10.1103/RevModPhys.65.611} {\bibfield
  {journal} {\bibinfo  {journal} {Rev. Mod. Phys.}\ }\textbf {\bibinfo {volume}
  {65}},\ \bibinfo {pages} {611} (\bibinfo {year} {1993})}\BibitemShut
  {NoStop}%
\bibitem [{\citenamefont {Harris}\ and\ \citenamefont {Jones}(1974)}]{Jones}%
  \BibitemOpen
  \bibfield  {author} {\bibinfo {author} {\bibfnamefont {J.}~\bibnamefont
  {Harris}}\ and\ \bibinfo {author} {\bibfnamefont {R.~O.}\ \bibnamefont
  {Jones}},\ }\href {http://stacks.iop.org/0305-4608/4/i=8/a=013} {\bibfield
  {journal} {\bibinfo  {journal} {J. Phys. F: Met. Phys.}\ }\textbf {\bibinfo
  {volume} {4}},\ \bibinfo {pages} {1170} (\bibinfo {year} {1974})}\BibitemShut
  {NoStop}%
\bibitem [{\citenamefont {Langreth}\ and\ \citenamefont
  {Perdew}(1975)}]{Langp1}%
  \BibitemOpen
  \bibfield  {author} {\bibinfo {author} {\bibfnamefont {D.}~\bibnamefont
  {Langreth}}\ and\ \bibinfo {author} {\bibfnamefont {J.}~\bibnamefont
  {Perdew}},\ }\href {\doibase http://dx.doi.org/10.1016/0038-1098(75)90618-3}
  {\bibfield  {journal} {\bibinfo  {journal} {Solid State Commun.}\ }\textbf
  {\bibinfo {volume} {17}},\ \bibinfo {pages} {1425 } (\bibinfo {year}
  {1975})}\BibitemShut {NoStop}%
\bibitem [{\citenamefont {Gunnarsson}\ and\ \citenamefont
  {Lundqvist}(1977)}]{Gun2}%
  \BibitemOpen
  \bibfield  {author} {\bibinfo {author} {\bibfnamefont {O.}~\bibnamefont
  {Gunnarsson}}\ and\ \bibinfo {author} {\bibfnamefont {B.~I.}\ \bibnamefont
  {Lundqvist}},\ }\href
  {http://journals.aps.org/prb/pdf/10.1103/PhysRevB.15.6006.3} {\bibfield
  {journal} {\bibinfo  {journal} {Phys. Rev. B}\ }\textbf {\bibinfo {volume}
  {15}},\ \bibinfo {pages} {6006} (\bibinfo {year} {1977})}\BibitemShut
  {NoStop}%
\bibitem [{\citenamefont {Langreth}\ and\ \citenamefont
  {Perdew}(1977)}]{Langp2}%
  \BibitemOpen
  \bibfield  {author} {\bibinfo {author} {\bibfnamefont {D.~C.}\ \bibnamefont
  {Langreth}}\ and\ \bibinfo {author} {\bibfnamefont {J.~P.}\ \bibnamefont
  {Perdew}},\ }\href {http://journals.aps.org/prb/pdf/10.1103/PhysRevB.15.2884}
  {\bibfield  {journal} {\bibinfo  {journal} {Phys. Rev. B}\ }\textbf {\bibinfo
  {volume} {15}},\ \bibinfo {pages} {2884} (\bibinfo {year}
  {1977})}\BibitemShut {NoStop}%
\bibitem [{\citenamefont {Frydel}\ \emph {et~al.}(2000)\citenamefont {Frydel},
  \citenamefont {Terilla},\ and\ \citenamefont {Burke}}]{Frydb}%
  \BibitemOpen
  \bibfield  {author} {\bibinfo {author} {\bibfnamefont {D.}~\bibnamefont
  {Frydel}}, \bibinfo {author} {\bibfnamefont {W.~M.}\ \bibnamefont {Terilla}},
  \ and\ \bibinfo {author} {\bibfnamefont {K.}~\bibnamefont {Burke}},\ }\href
  {\doibase 10.1063/1.481099} {\bibfield  {journal} {\bibinfo  {journal} {J.
  Chem. Phys.}\ }\textbf {\bibinfo {volume} {112}},\ \bibinfo {pages} {5292}
  (\bibinfo {year} {2000})}\BibitemShut {NoStop}%
\bibitem [{\citenamefont {Levy}\ and\ \citenamefont {Perdew}(1985)}]{LP85}%
  \BibitemOpen
  \bibfield  {author} {\bibinfo {author} {\bibfnamefont {M.}~\bibnamefont
  {Levy}}\ and\ \bibinfo {author} {\bibfnamefont {J.~P.}\ \bibnamefont
  {Perdew}},\ }\href {\doibase 10.1103/PhysRevA.32.2010} {\bibfield  {journal}
  {\bibinfo  {journal} {Phys. Rev. A}\ }\textbf {\bibinfo {volume} {32}},\
  \bibinfo {pages} {2010} (\bibinfo {year} {1985})}\BibitemShut {NoStop}%
\bibitem [{\citenamefont {Umrigar}\ and\ \citenamefont {Gonze}(1994)}]{Cyru}%
  \BibitemOpen
  \bibfield  {author} {\bibinfo {author} {\bibfnamefont {C.~J.~.}\ \bibnamefont
  {Umrigar}}\ and\ \bibinfo {author} {\bibfnamefont {X.}~\bibnamefont
  {Gonze}},\ }\href {http://journals.aps.org/pra/pdf/10.1103/PhysRevA.50.3827}
  {\bibfield  {journal} {\bibinfo  {journal} {Phys. Rev. A}\ }\textbf {\bibinfo
  {volume} {50}},\ \bibinfo {pages} {3827} (\bibinfo {year}
  {1994})}\BibitemShut {NoStop}%
\bibitem [{\citenamefont {Teale}\ \emph {et~al.}(2009)\citenamefont {Teale},
  \citenamefont {Coriani},\ and\ \citenamefont {Helgaker}}]{Teal2}%
  \BibitemOpen
  \bibfield  {author} {\bibinfo {author} {\bibfnamefont {A.~M.}\ \bibnamefont
  {Teale}}, \bibinfo {author} {\bibfnamefont {S.}~\bibnamefont {Coriani}}, \
  and\ \bibinfo {author} {\bibfnamefont {T.}~\bibnamefont {Helgaker}},\
  }\href@noop {} {\bibfield  {journal} {\bibinfo  {journal} {J. Chem. Phys.}\
  }\textbf {\bibinfo {volume} {130}},\ \bibinfo {pages} {104111} (\bibinfo
  {year} {2009})}\BibitemShut {NoStop}%
\bibitem [{\citenamefont {Kais}\ \emph {et~al.}(1993)\citenamefont {Kais},
  \citenamefont {Hersbach},\ and\ \citenamefont {Handy}}]{Kais}%
  \BibitemOpen
  \bibfield  {author} {\bibinfo {author} {\bibfnamefont {S.}~\bibnamefont
  {Kais}}, \bibinfo {author} {\bibfnamefont {D.~R.}\ \bibnamefont {Hersbach}},
  \ and\ \bibinfo {author} {\bibfnamefont {N.~C.}\ \bibnamefont {Handy}},\
  }\href {http://aip.scitation.org/doi/pdf/10.1063/1.465765} {\bibfield
  {journal} {\bibinfo  {journal} {J. Chem. Phys.}\ }\textbf {\bibinfo {volume}
  {99}},\ \bibinfo {pages} {417} (\bibinfo {year} {1993})}\BibitemShut
  {NoStop}%
\bibitem [{\citenamefont {Ernzerhof}(1996)}]{Ernz}%
  \BibitemOpen
  \bibfield  {author} {\bibinfo {author} {\bibfnamefont {M.}~\bibnamefont
  {Ernzerhof}},\ }\href {\doibase
  https://doi.org/10.1016/S0009-2614(96)01225-0} {\bibfield  {journal}
  {\bibinfo  {journal} {Chem. Phys. Lett.}\ }\textbf {\bibinfo {volume}
  {263}},\ \bibinfo {pages} {499 } (\bibinfo {year} {1996})}\BibitemShut
  {NoStop}%
\bibitem [{\citenamefont {Colonna}\ and\ \citenamefont {Savin}(1999)}]{Colon}%
  \BibitemOpen
  \bibfield  {author} {\bibinfo {author} {\bibfnamefont {F.}~\bibnamefont
  {Colonna}}\ and\ \bibinfo {author} {\bibfnamefont {A.}~\bibnamefont
  {Savin}},\ }\href@noop {} {\bibfield  {journal} {\bibinfo  {journal} {J.
  Chem. Phys.}\ }\textbf {\bibinfo {volume} {110}},\ \bibinfo {pages} {2828}
  (\bibinfo {year} {1999})}\BibitemShut {NoStop}%
\bibitem [{\citenamefont {Cohen}\ \emph {et~al.}(2007)\citenamefont {Cohen},
  \citenamefont {Mori-Sanchez},\ and\ \citenamefont {Yang}}]{Cohen}%
  \BibitemOpen
  \bibfield  {author} {\bibinfo {author} {\bibfnamefont {A.~J.}\ \bibnamefont
  {Cohen}}, \bibinfo {author} {\bibfnamefont {P.}~\bibnamefont {Mori-Sanchez}},
  \ and\ \bibinfo {author} {\bibfnamefont {W.}~\bibnamefont {Yang}},\ }\href
  {http://aip.scitation.org/doi/pdf/10.1063/1.2749510} {\bibfield  {journal}
  {\bibinfo  {journal} {J. Chem. Phys.}\ }\textbf {\bibinfo {volume} {127}},\
  \bibinfo {pages} {034101} (\bibinfo {year} {2007})}\BibitemShut {NoStop}%
\bibitem [{\citenamefont {Arbuznikov}\ and\ \citenamefont
  {Kaupp}(2008)}]{Alexe}%
  \BibitemOpen
  \bibfield  {author} {\bibinfo {author} {\bibfnamefont {A.~V.}\ \bibnamefont
  {Arbuznikov}}\ and\ \bibinfo {author} {\bibfnamefont {M.}~\bibnamefont
  {Kaupp}},\ }\href {http://aip.scitation.org/doi/pdf/10.1063/1.2920196}
  {\bibfield  {journal} {\bibinfo  {journal} {J. Chem. Phys.}\ }\textbf
  {\bibinfo {volume} {128}},\ \bibinfo {pages} {214107} (\bibinfo {year}
  {2008})}\BibitemShut {NoStop}%
\bibitem [{\citenamefont {Teale}\ \emph {et~al.}(2010)\citenamefont {Teale},
  \citenamefont {Coriani},\ and\ \citenamefont {Helgaker}}]{Teal3}%
  \BibitemOpen
  \bibfield  {author} {\bibinfo {author} {\bibfnamefont {A.~M.}\ \bibnamefont
  {Teale}}, \bibinfo {author} {\bibfnamefont {S.}~\bibnamefont {Coriani}}, \
  and\ \bibinfo {author} {\bibfnamefont {T.}~\bibnamefont {Helgaker}},\ }\href
  {\doibase 10.1063/1.3380834} {\bibfield  {journal} {\bibinfo  {journal} {J.
  Chem. Phys.}\ }\textbf {\bibinfo {volume} {132}},\ \bibinfo {pages} {164115}
  (\bibinfo {year} {2010})}\BibitemShut {NoStop}%
\bibitem [{\citenamefont {Su}\ and\ \citenamefont {Xu}(2014)}]{Neil}%
  \BibitemOpen
  \bibfield  {author} {\bibinfo {author} {\bibfnamefont {N.~Q.}\ \bibnamefont
  {Su}}\ and\ \bibinfo {author} {\bibfnamefont {X.}~\bibnamefont {Xu}},\ }\href
  {\doibase 10.1063/1.4866457} {\bibfield  {journal} {\bibinfo  {journal} {J.
  Chem. Phys.}\ }\textbf {\bibinfo {volume} {140}},\ \bibinfo {pages} {18A512}
  (\bibinfo {year} {2014})}\BibitemShut {NoStop}%
\bibitem [{\citenamefont {Chakravorty}\ \emph {et~al.}(1993)\citenamefont
  {Chakravorty}, \citenamefont {Gwaltney}, \citenamefont {Davidson},
  \citenamefont {Parpia},\ and\ \citenamefont {Fischer}}]{Shubh}%
  \BibitemOpen
  \bibfield  {author} {\bibinfo {author} {\bibfnamefont {S.~J.}\ \bibnamefont
  {Chakravorty}}, \bibinfo {author} {\bibfnamefont {S.~R.}\ \bibnamefont
  {Gwaltney}}, \bibinfo {author} {\bibfnamefont {E.~R.}\ \bibnamefont
  {Davidson}}, \bibinfo {author} {\bibfnamefont {F.~A.}\ \bibnamefont
  {Parpia}}, \ and\ \bibinfo {author} {\bibfnamefont {C.~F.}\ \bibnamefont
  {Fischer}},\ }\href {\doibase 10.1103/PhysRevA.47.3649} {\bibfield  {journal}
  {\bibinfo  {journal} {Phys. Rev. A}\ }\textbf {\bibinfo {volume} {47}},\
  \bibinfo {pages} {3649} (\bibinfo {year} {1993})}\BibitemShut {NoStop}%
\bibitem [{\citenamefont {McCarthy}\ and\ \citenamefont
  {Thakkar}(2011)}]{Shane}%
  \BibitemOpen
  \bibfield  {author} {\bibinfo {author} {\bibfnamefont {S.~P.}\ \bibnamefont
  {McCarthy}}\ and\ \bibinfo {author} {\bibfnamefont {A.~J.}\ \bibnamefont
  {Thakkar}},\ }\href {\doibase 10.1063/1.3547262} {\bibfield  {journal}
  {\bibinfo  {journal} {J. Chem. Phys.}\ }\textbf {\bibinfo {volume} {134}},\
  \bibinfo {pages} {044102} (\bibinfo {year} {2011})}\BibitemShut {NoStop}%
\bibitem [{\citenamefont {Vosko}\ \emph {et~al.}(1980)\citenamefont {Vosko},
  \citenamefont {Wilk},\ and\ \citenamefont {Nusair}}]{Vwn}%
  \BibitemOpen
  \bibfield  {author} {\bibinfo {author} {\bibfnamefont {S.~H.}\ \bibnamefont
  {Vosko}}, \bibinfo {author} {\bibfnamefont {L.}~\bibnamefont {Wilk}}, \ and\
  \bibinfo {author} {\bibfnamefont {M.}~\bibnamefont {Nusair}},\ }\href
  {http://www.nrcresearchpress.com/doi/pdf/10.1139/p80-159} {\bibfield
  {journal} {\bibinfo  {journal} {Can. J. Phys.}\ }\textbf {\bibinfo {volume}
  {58}},\ \bibinfo {pages} {1200} (\bibinfo {year} {1980})}\BibitemShut
  {NoStop}%
\bibitem [{\citenamefont {Chachiyo}(2016)}]{Teepa}%
  \BibitemOpen
  \bibfield  {author} {\bibinfo {author} {\bibfnamefont {T.}~\bibnamefont
  {Chachiyo}},\ }\href {\doibase 10.1063/1.4958669} {\bibfield  {journal}
  {\bibinfo  {journal} {J. Chem. Phys.}\ }\textbf {\bibinfo {volume} {145}},\
  \bibinfo {pages} {021101} (\bibinfo {year} {2016})}\BibitemShut {NoStop}%
\bibitem [{\citenamefont {Sottile}\ and\ \citenamefont
  {Ballone}(2001)}]{Sottile_PRB.64.045105}%
  \BibitemOpen
  \bibfield  {author} {\bibinfo {author} {\bibfnamefont {F.}~\bibnamefont
  {Sottile}}\ and\ \bibinfo {author} {\bibfnamefont {P.}~\bibnamefont
  {Ballone}},\ }\href {\doibase 10.1103/PhysRevB.64.045105} {\bibfield
  {journal} {\bibinfo  {journal} {Phys. Rev. B}\ }\textbf {\bibinfo {volume}
  {64}},\ \bibinfo {pages} {045105} (\bibinfo {year} {2001})}\BibitemShut
  {NoStop}%
\bibitem [{\citenamefont {Perdew}\ and\ \citenamefont {Wang}(1992)}]{Wang}%
  \BibitemOpen
  \bibfield  {author} {\bibinfo {author} {\bibfnamefont {J.~P.}\ \bibnamefont
  {Perdew}}\ and\ \bibinfo {author} {\bibfnamefont {Y.}~\bibnamefont {Wang}},\
  }\href {\doibase 10.1103/PhysRevB.45.13244} {\bibfield  {journal} {\bibinfo
  {journal} {Phys. Rev. B}\ }\textbf {\bibinfo {volume} {45}},\ \bibinfo
  {pages} {13244} (\bibinfo {year} {1992})}\BibitemShut {NoStop}%
\bibitem [{\citenamefont {Chauhan}\ and\ \citenamefont {Harbola}(2015)}]{Rabi}%
  \BibitemOpen
  \bibfield  {author} {\bibinfo {author} {\bibfnamefont {R.~S.}\ \bibnamefont
  {Chauhan}}\ and\ \bibinfo {author} {\bibfnamefont {M.~K.}\ \bibnamefont
  {Harbola}},\ }\href
  {http://ac.els-cdn.com/S0009261415007290/1-s2.0-S0009261415007290-main.pdf?_tid=1b53f778-daf9-11e6-b6a4-00000aab0f27&acdnat=1484467592_efe05321028d792b341cccd76f553746}
  {\bibfield  {journal} {\bibinfo  {journal} {Chem. Phys. Lett.}\ }\textbf
  {\bibinfo {volume} {639}},\ \bibinfo {pages} {248} (\bibinfo {year}
  {2015})}\BibitemShut {NoStop}%
\bibitem [{\citenamefont {Taut}(1993)}]{Taut}%
  \BibitemOpen
  \bibfield  {author} {\bibinfo {author} {\bibfnamefont {M.}~\bibnamefont
  {Taut}},\ }\href {http://journals.aps.org/pra/pdf/10.1103/PhysRevA.48.3561}
  {\bibfield  {journal} {\bibinfo  {journal} {Phys. Rev. A}\ }\textbf {\bibinfo
  {volume} {48}},\ \bibinfo {pages} {3561} (\bibinfo {year}
  {1993})}\BibitemShut {NoStop}%
\bibitem [{\citenamefont {Chauhan}\ and\ \citenamefont
  {Harbola}(2017)}]{Rabi2}%
  \BibitemOpen
  \bibfield  {author} {\bibinfo {author} {\bibfnamefont {R.~S.}\ \bibnamefont
  {Chauhan}}\ and\ \bibinfo {author} {\bibfnamefont {M.~K.}\ \bibnamefont
  {Harbola}},\ }\href {\doibase 10.1002/qua.25344} {\bibfield  {journal}
  {\bibinfo  {journal} {Int. J. Quantum Chem.}\ }\textbf {\bibinfo {volume}
  {117}},\ \bibinfo {pages} {25344} (\bibinfo {year} {2017})}\BibitemShut
  {NoStop}%
\bibitem [{\citenamefont {Sun}\ \emph {et~al.}(2016)\citenamefont {Sun},
  \citenamefont {Perdew}, \citenamefont {Yang},\ and\ \citenamefont
  {Peng}}]{Sun_16}%
  \BibitemOpen
  \bibfield  {author} {\bibinfo {author} {\bibfnamefont {J.}~\bibnamefont
  {Sun}}, \bibinfo {author} {\bibfnamefont {J.~P.}\ \bibnamefont {Perdew}},
  \bibinfo {author} {\bibfnamefont {Z.}~\bibnamefont {Yang}}, \ and\ \bibinfo
  {author} {\bibfnamefont {H.}~\bibnamefont {Peng}},\ }\href {\doibase
  10.1063/1.4950845} {\bibfield  {journal} {\bibinfo  {journal} {J. Chem.
  Phys.}\ }\textbf {\bibinfo {volume} {144}},\ \bibinfo {pages} {191101}
  (\bibinfo {year} {2016})}\BibitemShut {NoStop}%
\bibitem [{\citenamefont {Fischer}(1987)}]{hf86}%
  \BibitemOpen
  \bibfield  {author} {\bibinfo {author} {\bibfnamefont {C.~F.}\ \bibnamefont
  {Fischer}},\ }\href {\doibase http://dx.doi.org/10.1016/0010-4655(87)90053-1}
  {\bibfield  {journal} {\bibinfo  {journal} {Comput. Phys. Commun.}\ }\textbf
  {\bibinfo {volume} {43}},\ \bibinfo {pages} {355 } (\bibinfo {year}
  {1987})}\BibitemShut {NoStop}%
\bibitem [{\citenamefont {Ashcroft}\ and\ \citenamefont {Mermin}(1976)}]{Ashc}%
  \BibitemOpen
  \bibfield  {author} {\bibinfo {author} {\bibfnamefont {N.}~\bibnamefont
  {Ashcroft}}\ and\ \bibinfo {author} {\bibfnamefont {N.}~\bibnamefont
  {Mermin}},\ }\href@noop {} {\emph {\bibinfo {title} {{Solid State
  Physics}}}}\ (\bibinfo  {publisher} {Saunders College},\ \bibinfo {address}
  {Philadelphia},\ \bibinfo {year} {1976})\BibitemShut {NoStop}%
\bibitem [{\citenamefont {Harbola}\ and\ \citenamefont {Sahni}(1989)}]{Harsa}%
  \BibitemOpen
  \bibfield  {author} {\bibinfo {author} {\bibfnamefont {M.~K.}\ \bibnamefont
  {Harbola}}\ and\ \bibinfo {author} {\bibfnamefont {V.}~\bibnamefont
  {Sahni}},\ }\href {\doibase 10.1103/PhysRevLett.62.489} {\bibfield  {journal}
  {\bibinfo  {journal} {Phys. Rev. Lett.}\ }\textbf {\bibinfo {volume} {62}},\
  \bibinfo {pages} {489} (\bibinfo {year} {1989})}\BibitemShut {NoStop}%
\bibitem [{\citenamefont {Sahni}(2016)}]{QDFT}%
  \BibitemOpen
  \bibfield  {author} {\bibinfo {author} {\bibfnamefont {V.}~\bibnamefont
  {Sahni}},\ }\href
  {https://link.springer.com/book/10.1007%2F978-3-662-49842-2} {\emph {\bibinfo
  {title} {Quantal Density Functional Theory}}}\ (\bibinfo  {publisher}
  {Springer-Verlag Berlin Heidelberg},\ \bibinfo {year} {2016})\BibitemShut
  {NoStop}%
\bibitem [{\citenamefont {Colle}\ and\ \citenamefont
  {Salvetti}(1975)}]{Colle1975}%
  \BibitemOpen
  \bibfield  {author} {\bibinfo {author} {\bibfnamefont {R.}~\bibnamefont
  {Colle}}\ and\ \bibinfo {author} {\bibfnamefont {O.}~\bibnamefont
  {Salvetti}},\ }\href {\doibase 10.1007/BF01028401} {\bibfield  {journal}
  {\bibinfo  {journal} {Theor. Chim. Acta}\ }\textbf {\bibinfo {volume} {37}},\
  \bibinfo {pages} {329} (\bibinfo {year} {1975})}\BibitemShut {NoStop}%
\bibitem [{\citenamefont {Singh}\ \emph {et~al.}(1999)\citenamefont {Singh},
  \citenamefont {Massa},\ and\ \citenamefont {Sahni}}]{Ranbir}%
  \BibitemOpen
  \bibfield  {author} {\bibinfo {author} {\bibfnamefont {R.}~\bibnamefont
  {Singh}}, \bibinfo {author} {\bibfnamefont {L.}~\bibnamefont {Massa}}, \ and\
  \bibinfo {author} {\bibfnamefont {V.}~\bibnamefont {Sahni}},\ }\href
  {\doibase 10.1103/PhysRevA.60.4135} {\bibfield  {journal} {\bibinfo
  {journal} {Phys. Rev. A}\ }\textbf {\bibinfo {volume} {60}},\ \bibinfo
  {pages} {4135} (\bibinfo {year} {1999})}\BibitemShut {NoStop}%
\bibitem [{\citenamefont {Tao}\ \emph {et~al.}(2001)\citenamefont {Tao},
  \citenamefont {Gori-Giorgi}, \citenamefont {Perdew},\ and\ \citenamefont
  {McWeeny}}]{Tao63}%
  \BibitemOpen
  \bibfield  {author} {\bibinfo {author} {\bibfnamefont {J.}~\bibnamefont
  {Tao}}, \bibinfo {author} {\bibfnamefont {P.}~\bibnamefont {Gori-Giorgi}},
  \bibinfo {author} {\bibfnamefont {J.~P.}\ \bibnamefont {Perdew}}, \ and\
  \bibinfo {author} {\bibfnamefont {R.}~\bibnamefont {McWeeny}},\ }\href
  {\doibase 10.1103/PhysRevA.63.032513} {\bibfield  {journal} {\bibinfo
  {journal} {Phys. Rev. A}\ }\textbf {\bibinfo {volume} {63}},\ \bibinfo
  {pages} {032513} (\bibinfo {year} {2001})}\BibitemShut {NoStop}%
\end{thebibliography}
%

\end{document}